\documentclass[sigconf,10pt]{acmart}

\usepackage{booktabs}
\usepackage{multicol}
\usepackage{tabularx}
\usepackage{tabulary}
\usepackage{enumitem}
\usepackage{multirow} % for merging rows
\usepackage{multicol} % for merging columns
\usepackage{adjustbox} % for rotating tables
\usepackage{caption} 
\usepackage{url}

\AtBeginDocument{%
  }

\acmYear{2024}\copyrightyear{2024}
\acmConference[ACM MobiCom '24]{International Conference On Mobile Computing And Networking}{September 30--October 4, 2024}{Washington D.C., DC, USA}
\acmBooktitle{International Conference On Mobile Computing And Networking (ACM MobiCom '24), September 30--October 4, 2024, Washington D.C., DC, USA}
\acmDOI{10.1145/3636534.3649371}
\acmISBN{979-8-4007-0489-5/24/09}

\renewcommand{\arraystretch}{1.3}

\begin{document}

\title{Deciphering the Enigma of Satellite Computing with COTS Devices: Measurement and Analysis}


\author{Ruolin Xing$^{\blacklozenge}$, Mengwei Xu$^{\blacklozenge}$, Ao Zhou$^{\blacklozenge}$, Qing Li$^{\Diamond}$ \\ Yiran Zhang$^{\blacklozenge}$, Feng Qian$^{\blacktriangle}$, Shangguang Wang$^{\blacklozenge}$}
\affiliation{%
  \institution{$^{\blacklozenge}$Beijing University of Posts and Telecommunications, $^{\Diamond}$Peking University, $^{\blacktriangle}$University of Southern California}
  \country{}
}

\renewcommand{\shorttitle}{Deciphering the Enigma of Satellite Computing with COTS Devices}
\renewcommand{\shortauthors}{Xing et al.}


\begin{abstract}
In the wake of the rapid deployment of large-scale low-Earth orbit satellite constellations, exploiting the full computing potential of Commercial Off-The-Shelf (COTS) devices in these environments has become a pressing issue.
However, understanding this problem is far from straightforward due to the inherent differences between the terrestrial infrastructure and the satellite platform in space.
In this paper, we take an important step towards closing this knowledge gap by presenting the first measurement study on the thermal control, power management, and performance of COTS computing devices on satellites.
Our measurements reveal that the satellite platform and COTS computing devices significantly interplay in terms of the temperature and energy, forming the main constraints on satellite computing.
Further, we analyze the critical factors that shape the characteristics of onboard COTS computing devices.
We provide guidelines for future research on optimizing the use of such devices for computing purposes.
Finally, we have released the datasets to facilitate further study in satellite computing.
\end{abstract}

\begin{CCSXML}
<ccs2012>
<concept>
<concept_id>10003033.10003079.10011704</concept_id>
<concept_desc>Networks~Network measurement</concept_desc>
<concept_significance>300</concept_significance>
</concept>
<concept>
<concept_id>10010583.10010737.10010749</concept_id>
<concept_desc>Hardware~Testing with distributed and parallel systems</concept_desc>
<concept_significance>300</concept_significance>
</concept>
</ccs2012>
\end{CCSXML}

\ccsdesc[300]{Networks~Network measurement}
\ccsdesc[300]{Hardware~Testing with distributed and parallel systems}

\keywords{Satellite Networking, Earth Observation, Satellite Computing, Network Measurement, Dataset}

\maketitle

\newcommand{\satlt}{BUPT-1}

\section{Introduction}

The resurrection of aerospace technologies is making low-Earth orbit (LEO) satellites a promising frontier in mobile edge computing \cite{DBLP:conf/asplos/DenbyL20, DBLP:conf/asplos/DenbyCCLN23, DBLP:conf/wmcsa/KothariLL20, DBLP:conf/hotnets/BhattacherjeeKL20}.
The Commercial Off-The-Shelf (COTS) computing devices are becoming core units of in-orbit computing. They offer the unparalleled advantage of reusing advanced embedded systems deployed on Earth, thereby facilitating rapid development and deployment of computational capabilities. Several initiatives have successfully employed COTS computing devices to boost LEO satellite computational power. For instance, ESA's PhiSat-1 \cite{giuffrida2021varphi} using Intel's specialized image processing chip to achieving high-speed, low-power in-orbit image data processing, Cubesat Missions \cite{gas2022gaspacs,maskey2020cubesatnet}, and even more commercial projects \cite{safyan2020planet,lemur22023,exospace2023,ubotica2023,arkedgespace2023}.

Utilizing COTS devices in space for computing purposes is challenging due to the harsh, distinct space environment compared to those on Earth.
This is primarily manifested in two aspects: (1) Thermal Control: The vacuum of space prevents heat dissipation through air convection \cite{miao2021design}. The frequent alternation between daylight and eclipse zones leads to significant variations in the thermal conditions external to the satellite systems \cite{tachikawa2022advanced}. (2) Energy Management: Satellites can only capture and store a very limited amount of solar energy \cite{knap2020review}. The energy is subject to internal contention within the satellite systems \cite{white2017commercial}, necessitating prioritization for the normal operation of the satellite platform \cite{aslanov2021chaos}. Additionally, radiation effects can induce damage or anomalies in COTS semiconductor devices. Such effects are generally not pronounced at LEO around 500 km. The impact of radiation effects is briefly discussed in \S\ref{subsec:discus}.

Within the satellite industry, pre-launch methods \cite{JAXA-JERG-2-310, JAXA-JERG-2-214, wertz1999space, nasa2022sotasat} 
are used to address these issues by incorporating \textit{sufficient design margins}: 
(1) For thermal control, the design \cite{miao2021design} of thermal structures is informed by results from thermal analysis \cite{JAXA-JERG-2-310}, with appropriate thermal materials \cite{tachikawa2022advanced} selected to facilitate insulation, conduction, and dissipation processes, culminating in tests such as thermal vacuum experiments. Across these critical steps, margins are incorporated to ensure that, the thermal load on various satellite components does not exceed threshold values under normal conditions. For instance, if the design temperature range for the satellite interior is -10 to 30\textdegree C, the active and passive thermal management systems typically limit temperatures to a range of 0 to 20\textdegree C.
(2) Energy management primarily involves the acquisition, storage, and distribution of energy, corresponding to the solar cells, batteries, and power control/distribution boards, respectively. These key components undergo simulation, selection, design, and testing to ensure their functionality\cite{JAXA-JERG-2-214, ribah2019power, li2021optimal}. To accommodate the satellite's overall power consumption budget, energy management systems also incorporate margins. For example, the remaining battery capacity in satellites is usually maintained above 70\% to ensure sufficient energy availability for unexpected demands, while also mitigating battery degradation.

Despite the rigorous processes to ensure operational integrity, a pivotal question remains unanswered: to what extent do thermal control and energy management impact the \textit{computing capability} and \textit{reliability} of COTS computing devices in actual satellite systems? Furthermore, how can computing tasks be scheduled to both comply with the constraints of the satellite platform and maximize computing capabilities?

To this end, the satellite community has developed mathematical models and simulation tools for analyzing temperature \cite{JAXA-JERG-2-310, miao2021design, tachikawa2022advanced, farrahi2017simplified, chen2023thermal} and energy \cite{JAXA-JERG-2-214, ribah2019power, ehren2019energy, li2021optimal} variations mainly in the pre-launch stage, aiding in the theoretical assessment of temperature and energy constraints. The computer science community has devised simulation platforms \cite{lai2020starperf, lai2023starrynet, DBLP:conf/middleware/PfandzelterB22} tailored for satellite networks and in-orbit computing, capable of conducting pure or hardware-in-the-loop simulations of satellite computing workflows. Building on this foundation, \textit{real-world data} would enable the synergistic integration of both strands of work, facilitating simulations of satellite computing that mirror authentic conditions and allowing for the effective scheduling of actual computing tasks on satellites.

However, a genuine in-orbit testing of COTS computing devices presents a plethora of challenges that extend beyond mere technical considerations.
First, the construction of a real satellite system is an expensive and intricate process.
Even for a small LEO satellite, the costs can be ranging from tens to hundreds of millions of dollars \cite{nanoavionics2022}.
The "one-shot" nature of satellite make it impossible to maintain the system \cite{NASA_CubeSat101}.
Second, COTS computing devices are exposed to a highly dynamic environment, making experimental design challenging when trying to eliminate the irrelevant factors \cite{nasa2022sotasat}. For instance, the eclipse and daylight periods, the thermal structure of the payloads, differences in payloads, and variations in work loads can all influence temperature and energy consumption measurements.
These multiple factors converge, creating a complex system landscape with various variables and extensive experimentation. This complexity makes direct comparisons and clear conclusions challenging.

To overcome these challenges, we invest two years and over one million dollars in preparation and testing, creating a realistic satellite system. The satellite has been operating normally in orbit for over 1 year, completing more than 6000 circuits around the Earth. We have successfully conducted various experiments on typical COTS computing devices, amounting to over 1000 hours of investigative work. Simultaneously, we develop a terrestrial testbed, replicating the satellite structure for COTS computing devices in a 1:1 ratio to minimize the impact of unrelated factors. We design a series of experiments for key factors that affect temperature and energy, classifying and contrasting each major influencing factor. After 6-month experimental data collection, we amass 10,000,000+ lines of telemetry information and over 22GB of trimmed text data. The findings from these experiments can be summarized as follows:

$\bullet$ \textbf{Temperature---the Critical Bottleneck for In-orbit Computing.} The heating of COTS device chips on satellites may lead to frequency throttling, resulting in an up to 10\% reduction in computational performance. A computation task lasting 10 hours, operating at approximately 9 Watts, causes the surface temperature to rise beyond the operational limit (e.g. 30℃ in our case) \cite{wertz1999space}, leading to instability. This phenomenon can be attributed to the passive heat dissipation mechanism\cite{nasa2022sotasat}. Limited by the satellite's volume and weight budget, active overheating control is generally unacceptable. In comparison, the terrestrial systems can leverage air convection and cooling devices to quickly stabilize temperature. The implication is a need for computational task planning in orbit to avert prolonged computation and prevent overheating.

$\bullet$ \textbf{The Influence of Eclipse and Daylight on Computing.} The eclipse or daylight periods can slightly affect payload chip and surface temperatures, with variations mostly within \( \pm 5\,^{\circ}\text{C} \). However, regardless of whether the cycle begins during the eclipse or daylight periods, the temperatures at the beginning and ending times remain roughly similar, with a difference of less than 2℃. This phenomenon results from the Sun being the primary external heat source, with the change in solar exposure affecting COTS device cooling \cite{nasa2022sotasat}. For long-duration satellite computing tasks, eclipse or daylight launch consideration is unnecessary, while short-duration computing tasks may benefit from launching in eclipse period. This contrasts with ground edge computing devices, where active temperature control renders environmental temperature changes insignificant.

$\bullet$ \textbf{Computing Dominance on Short-term Power Consumption.} Even over short periods (e.g. within a single day), the power consumption of COTS computing devices can have a significant impact on the battery's Depth of Discharge (DoD), potentially causing it to exceed 30\%, even 50\%. The satellite's steady-state power consumption is relatively stable, but an increase in the average DoD from 25\% to 30\% can reduce the battery's lifespan by approximately 25\% \cite{fellner2003lithium}, consequently affecting the satellite's operational longevity. Unlike terrestrial edge computing devices, where battery maintenance or replacement is feasible, the satellite computing tasks must softly control the maximum DoD within 30\% for short-term intensive computing.

$\bullet$ \textbf{Potential Energy Efficiency Bonus for Computing.} We observe a cyclical pattern in satellite energy collection over two distinct periods: one orbit around Earth and one natural day. But approximately 6\% of the converted solar energy remains unutilized. On the other hand, we note that there is a 30\% to 40\% energy consumption occupied by the computing tasks while the communication power consumption and others power consumption remain stable. This discrepancy arises from an imperfect alignment between the scheduling of satellite computations and the periodic energy collection \cite{wertz1999space}. When the battery reaches full capacity, the residual solar energy is not consumed by the computational tasks. Unlike typical terrestrial edge devices with on-demand energy supply, in-orbit computing  scheduling is better to consider these intricate cycles, optimizing the utilization of solar energy that cannot be stored.

The above results unequivocally establish that \textit{both temperature and energy are significant limiting factors that directly impede the capability and even reliability of onboard COTS computing devices}. In extreme conditions such as overheating, computing tasks, as well as COTS computing devices, may be subjected to an involuntary shutdown.
Furthermore, we identify surface temperature and DoD as the key metrics for quantitatively evaluating the constrains imposed by the satellite platform.
Given these insights, we advocate for the community to increase focus on the challenges of computing task scheduling under platform constraints to finally meet the goal of long-term stable execution of in-orbit computing tasks.
In summary, our main contributions are as follows:
\begin{list}{$\bullet$}{\leftmargin=1em \itemindent=0em \topsep=0.1em}
\item We conduct extensive measurements on a real satellite equipped with COTS computing devices, filling the gap in empirical evaluation of satellite computing.
\item We provide insightful guidance for future research work related to hardware-in-the-loop simulation or task scheduling in the context of satellite computing.
\item We have released the detailed datasets at \url{https://github.com/TiansuanConstellation/MobiCom24-SatelliteCOTS}.

\end{list}

\section{Measurement Methodology}

\newlength{\oldtextfloatsep}
\setlength{\oldtextfloatsep}{\textfloatsep}
\setlength{\textfloatsep}{5pt}
\captionsetup{skip=0pt, belowskip=-4pt}
\begin{figure}
\scalebox{0.47}{\includegraphics[width=\textwidth, trim={6cm 3cm 6cm 3.5cm}, clip]{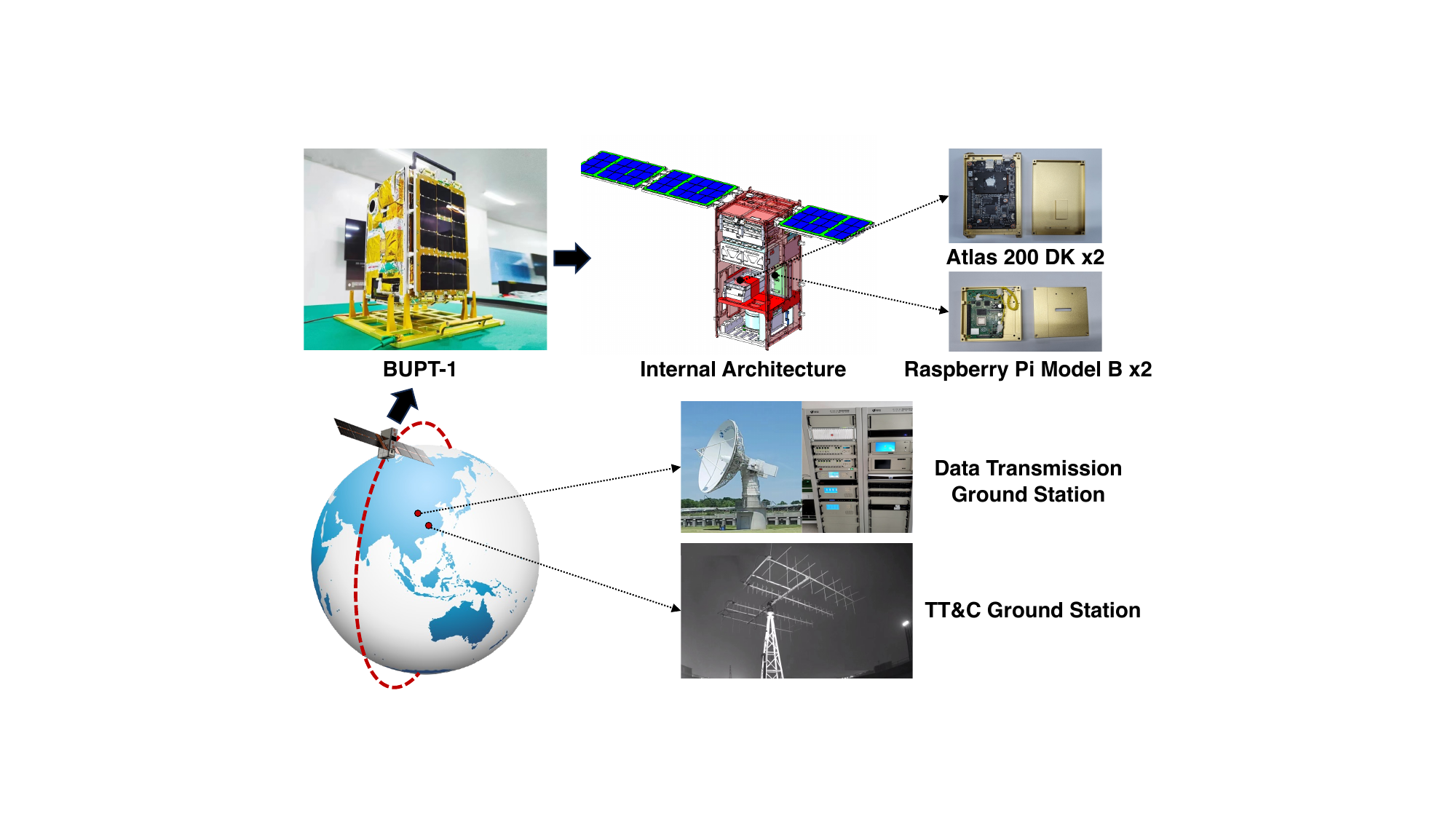}}
\caption{An overview of the measurement setup.}
\Description{An overview of the measurement setup.}
\label{fig:ov1}
\end{figure}
\captionsetup{skip=12pt, belowskip=0pt}

\subsection{Real-world Satellite System}
\label{subsec:oursat}

The \satlt\ is a 12U small satellite platform, with performance parameters slightly exceeding mainstream 6U and smaller CubeSats \cite{UCS}. The platform can provide robust support for COTS computing devices, i.e. the payloads. As indicated in Table~\ref{tab:bupt1-info}, it operates in a Sun-synchronous orbit, a common choice for experimental satellites \cite{swartwout2013first}.
Therefore, our experiments on \satlt\ regarding temperature control and power management can reflect the mutual interactions and impacts between satellite COTS computing devices and the satellite platform, without losing generality.

We use two ground stations to control the satellite(see Fig. \ref{fig:ov1}). The telemetry, tracking, and control (TT\&C) station enables loading command sequences onto the satellite for execution and controls. The data transmission station is responsible for receiving the telemetries\cite{CCSDS_132_0_B_3} and payload data generated by the satellite platform and COTS computing devices respectively.

\subsection{Onboard Devices and Experiments}
\label{subsec:assumption}

\textbf{Hardware.} Two types of COTS computing devices, Raspberry Pi 4B and Atlas 200 DK are selected for the experiments. The Raspberry Pi 4B, already employed in several successful in-orbit missions \cite{gas2022gaspacs, nepp2021processor}, has proven its feasibility and stability. Featuring an ARM Cortex-A72 chip with four cores, it is well-suited for general-purpose in-orbit processing such as photo control, image processing, and AI inference. The Atlas 200 DK integrates the Ascend 310 AI processor \cite{huawei_ascend_310} specifically for AI applications. This processor has built-in circuits capable of handling tasks like image and video encoding/decoding, AI classification, and AI inference. Furthermore, Atlas 200 DK offers four adjustable computational levels, i.e. Low, Mid, High, Full, corresponding to varying power consumption and computing performance.

As shown in Fig. \ref{fig:ov1}, we deploy two Raspberry Pi 4B and two Atlas 200 DK. These COTS computing devices were labeled for identification purposes, with the Raspberry Pi denoted as Pi-A and Pi-B, and the Atlas 200 DK denoted as Atlas-A and Atlas-B. Pi-A and Atlas-B were primarily utilized for most of the experiments, while the other two devices served as control groups for comparing.

\setlength{\textfloatsep}{\oldtextfloatsep}
\setlength{\textfloatsep}{0pt}
% \captionsetup{skip=3pt, belowskip=-16pt}
\begin{table}[tp]
\centering
\caption{Basic Information of \satlt}
\footnotesize % set font size to small
\renewcommand{\arraystretch}{0.95}
\begin{tabular}{p{3.5cm}p{4cm}}
\toprule
Parameter & Value \\
\midrule
ORCID & 55261 \\
Orbit & Sun-synchronous Orbit \\
Orbital Altitude & 487.607 km to 494.651 km \\
Orbital Inclination & 97.3710° \\
Mass & 17.44 kg \\
Volume & 434.5 mm $\times$ 346.6 mm $\times$ 340.8 mm \\
Internal Thermal Range & -10℃ to 30℃ \\
Solar Panel Size & 1199.3 mm $\times$ 868.4 mm $\times$ 423.4 mm \\
Battery Capacity & 115 Wh $\times$ 2 \\
TT\&C & 4.8 kbps uplink, 9.6 kbps downlink \\
Data Transmission & 1 Mbps uplink, 100 Mbps downlink \\
\bottomrule
\end{tabular}
\label{tab:bupt1-info}
\end{table}
% \captionsetup{skip=12pt, belowskip=0pt}
\setlength{\textfloatsep}{\oldtextfloatsep}

\textbf{Experiments.} On Raspberry Pi 4B, we deploy three kinds of experiments, i.e. stress tests, image segmentation, and image inference. The stress tests are divided into four groups, with each group respectively invoking 1 to 4 CPU cores (denoted as level 1 to 4). The image inference utilized 4 different models, i.e. SSD-MV1, YOLO-Fastest, YOLOv3, and YOLOv5-Lite. On the Atlas 200 DK, we also deploy experiments in three main categories. The first category is image encoding (denoted as 'jpege'). The second is image classification (denoted as 'imgcl'), utilizing the ResNet50 model. The third is object detection (denoted as 'od'), utilizing the YOLOv3 model for image or video stream inference. On Atlas, the inference tasks can be offloaded to the AI processor by 1 or 4 threads (denoted as 1T or 4T).

\textbf{Characteristics.} The satellite computing tasks generally have the following traits:
\begin{enumerate}[leftmargin=1.5em, itemindent=0em, topsep=0.1em]
\item The data generation rate far exceeds the processing capability \cite{nasa2015esds, doug2020teraspace}. This implies the strong motivations of satellite computing.
\item The tasks are stateless, meaning previous computations do not affect current ones. This is applicable to satellite-generated raw data, like images. Common image processing tasks like segmentation \cite{fernando2023towards}, classification \cite{maskey2020cubesatnet}, or object detection \cite{reichstein2019deep, kucik2021investigating} fit the stateless definition.
\item The computing tasks could tolerate a certain degree of latency. Low-latency processing is a key metric for satellite tasks \cite{DBLP:conf/sigcomm/VasishtSC21}, often demanding more computing resources in a short time. However, within the constraints of the satellite platform, the focus is not on real-time requirements but more on the computational volume \cite{DBLP:conf/asplos/DenbyL20}.
In this paper, we focus on the latter case, aiming at maximizing the processed data volume.
\end{enumerate}

\begin{table}[]
\centering
\caption{Collected Data in Telemetries}
\footnotesize
\renewcommand{\arraystretch}{0.95}
%\resizebox{0.96\columnwidth}{!}{%
\begin{tabular}{lll}
\toprule
Date Name                      & Unit  & Sample Rate \\
\midrule
Platform Total Voltage/Current & mV/mA & 1s          \\
Battery Voltage/Current        & mV/mA  & 4s\\
MPPT Input Voltage/Current     & mV/mA& 3s\\
MPPT Output Current    & mA &3s\\
Atlas-A/B Current        & mA & 1s\\
Pi-A/B Current        & mA & 1s\\
X-Band Transceiver Current  & mA & 1s\\
TT\&C Module Current & mA & 1s\\     
\midrule
Surface Temperature of Atlas-A/B       & \textdegree C & 4s\\
Surface Temperature of Pi-A/B      & \textdegree C & 4s\\
\bottomrule
\end{tabular}%
%}
\label{tab:tele-metrics}
\end{table}

\subsection{Data Collection and Processing}
\label{subsec:3-3}
\textbf{Payload and Telemetry Data.} The data collected from \satlt\ can be divided into payload and telemetry data. Payload data is generated by the onboard COTS computing devices. It includes measurements like chip temperature, CPU frequency, CPU usage per process, memory occupancy, image inference latency, etc. Telemetry data is produced by the satellite platform. It mainly encompasses categories of temperature, energy consumption, satellite attitude, orbit, and more. We list the essential data types in Table \ref{tab:tele-metrics}. From the table, the term \textit{MPPT} stands for maximum power point tracking, a technique employed in photovoltaic solar systems to optimize the power output.

\textbf{Data Collection.} The \satlt\ is successfully launched into orbit on January 15th. After approximately two months of stability testing, we conduct official experiments from March 22nd to July 30th, resulting in over 22GB of telemetry data and over 30GB of payload experiment data. The datasets are available at \url{https://github.com/TiansuanConstellation/MobiCom24-SatelliteCOTS}.

\textbf{Representativeness.} 
Influences introduced by irrelevant factors may destroy the representativeness of our study.
We make the following 3 efforts to eliminate irrelevant factors in terms of the thermal conditions, the time variations and the program correctness:
\begin{list}{$\bullet$}{\leftmargin=1em \itemindent=0em \topsep=0.1em}
\item We build a testbed on the ground for the two COTS computing devices with the completely same thermal alteration (see Fig. \ref{fig:ov1}).
\item We keep the time of the onboard COTS computing devices synchronized with the standard UTC time to avoid errors introduced by time variations.
\item We develop a program to generate identical command sequence both on satellites and the ground, minimizing the error introduced by the software to the greatest extent.
\end{list}

\captionsetup{skip=3pt, belowskip=-16pt}
\begin{figure}[pt]
\centering
\includegraphics[scale=0.255, trim={0 0.8cm 0 0}, clip]{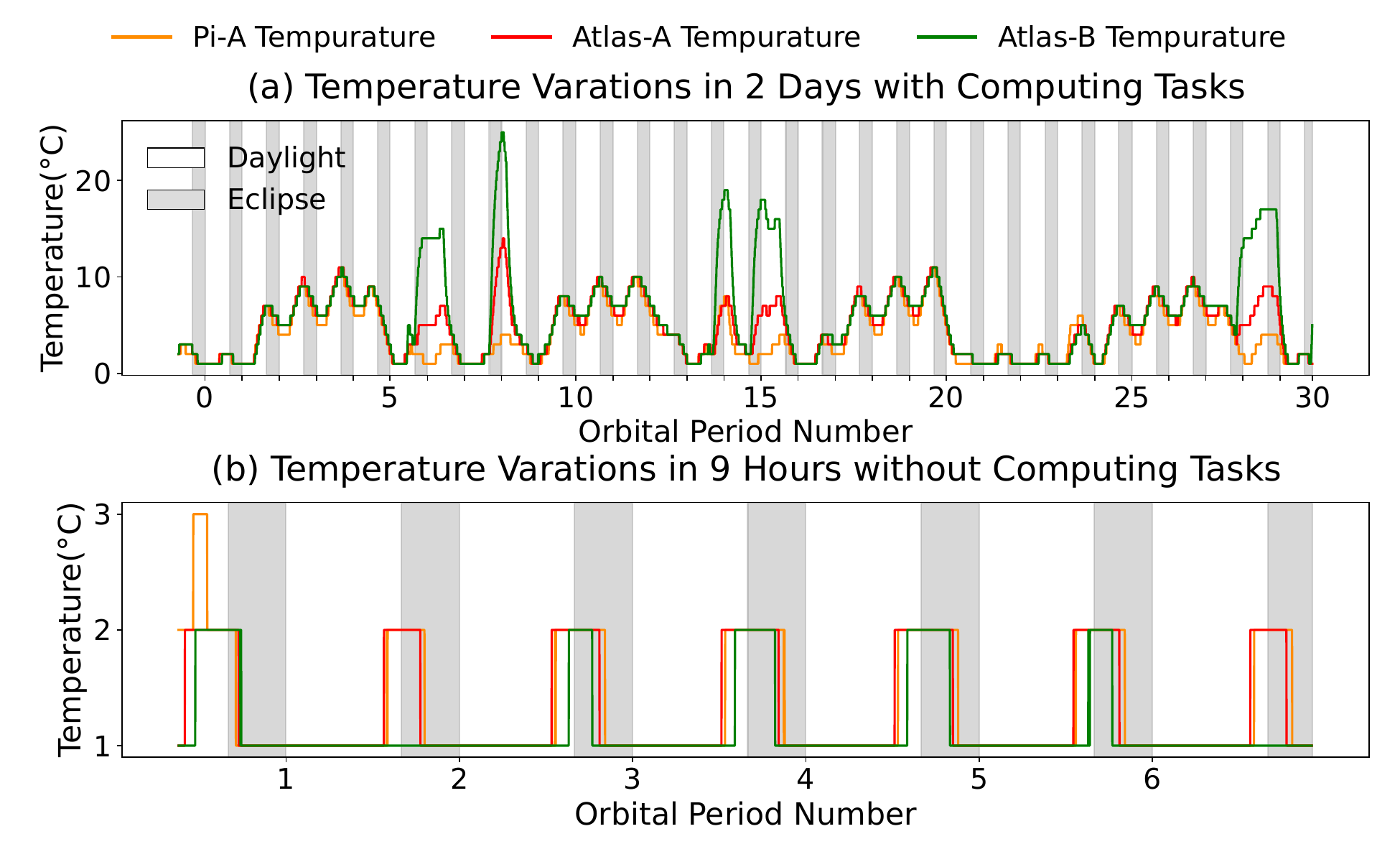}
\caption{Temperature Overview.}
\Description{Void.}
\label{fig:ov2}
\end{figure}
\captionsetup{skip=12pt, belowskip=0pt}

\section{Temperature Results}

\captionsetup{skip=2pt, belowskip=-16pt}
\begin{figure*}[pt]
\centering
\includegraphics[scale=0.39, trim={0 0 0 0}, clip]{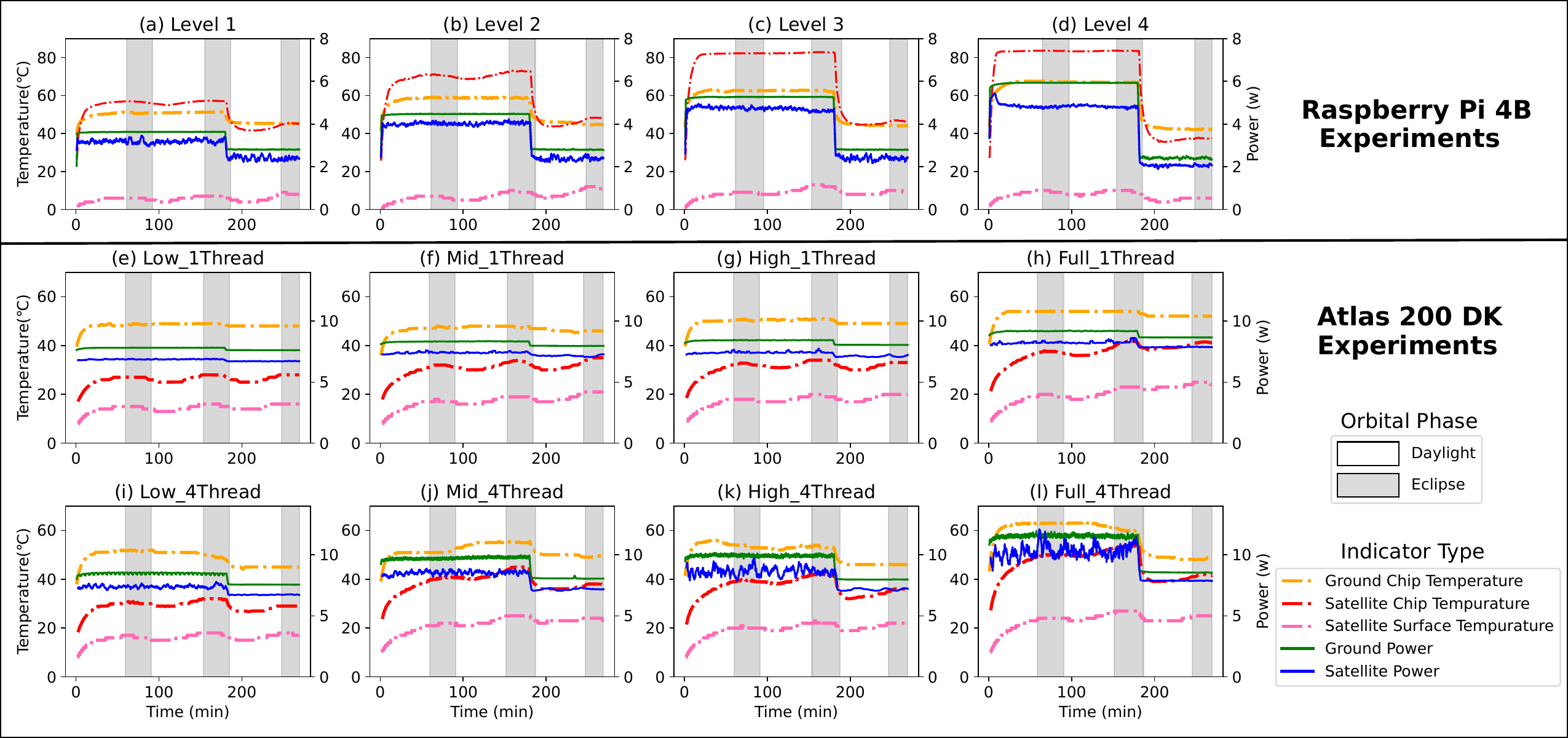}
\caption{Temperature and Power Variations for Atlas and Pi in Satellite-Terrestrial Environments.}
\Description{Void.}
\label{fig:tmp1}
\end{figure*}
\captionsetup{skip=12pt, belowskip=-46pt}

For satellites equipped with COTS computing devices, the typical temperature control process involves \cite{nasa2022sotasat, tachikawa2022advanced}: (1) heat generation from the computing tasks; (2) the heat transfer to a spreader through thermal pads and an aluminum alloy enclosure; (3) the subsequent conduction of heat from the spreader to the satellite's external structure for dissipation through radiation. Fig. \ref{fig:ov1} illustrates the thermal structure of the satellite, where the deep red area represents the heat spreader and the light red area indicates the satellite's external structure utilized for heat dissipation. Table \ref{tab:bupt1-info} presents the internal temperature design range for the satellite, set between -10 to 30\textdegree C. Therefore, the temperatures of the aluminum alloy enclosures and the heat spreader must not exceed this range, as it could potentially impact the satellite platform equipment and lead to overall system instability.

Uncontrolled temperature variations can have catastrophic consequences for a satellite. If the temperature surpasses its operational limit \cite{wertz1999space}, components may degrade or fail, leading to diminished performance or total system failure \cite{tachikawa2022advanced}. In extreme instances, this could culminate in irreparable damage to the spacecraft, underscoring the paramount importance of maintaining thermal regulation \cite{miao2021design}. Although COTS computing devices are subjected to thermal analysis, design, and testing in the pre-launch stage, there remains a potential risk for excessive internal satellite temperatures during prolonged durations under heavy workloads. This risk arises primarily from two factors:
\begin{enumerate}[leftmargin=1.5em, itemindent=0em, topsep=0.1em]
\item COTS computing devices and their accompanying cooling structures occupy a significant volume and weight on small satellites (see \S\ref{subsec:oursat}).
Under the passive cooling architecture, once the heat generation rate of the COTS computing devices surpass the heat dissipation rate of the satellite's external structure, the surface temperature will rapidly increase. Given the relatively low efficiency of radiative cooling, prolonged high-load computing can easily lead to overheating within the satellite's interior.

\item The thermal vacuum testing ensures that COTS computing devices can function at sufficiently high power in a vacuum environment, but only for a relatively short duration. The design of thermal isolation can largely isolate the heat exchange between COTS computing devices and the external environment as well as critical internal components of the satellite \cite{miao2021design}.
However, these testings are unable to guarantee the safety of the real satellite system once the internal temperature exceeds the designed safe range.
\end{enumerate}

To study the actual impact of COTS computing devices on temperature, we define the \textit{surface temperature} to indicate the temperature of the aluminum alloy enclosure and the heat spreader.
We define the unified temperature metric for the two components because the heat spreader and enclosure rapidly achieve thermal equilibrium.
The surface temperature should be within the range of -10 to 30\textdegree C.
Furthermore, the rise in enclosure temperature originates from the increase in the chip temperature of the COTS computing devices.

Consequently, we also emphasize the \textit{chip temperature} as the indicator of heat generation.
The limitations on chip temperature are specified in their respective documentation \cite{raspberry_pi_4, huawei_atlas_200}.
Finally, we use \textit{external temperature} to indicate the temperature of the satellite's external structure.

Additionally, the alternation between daylight and eclipse zones have an direct impact on the external temperature. Recall that the aluminum alloy enclosure of COTS computing devices ultimately dissipates heat through radiation via the satellite external structures (see Fig. \ref{fig:ov1}). And variations in external temperature can affect the cooling efficiency. Furthermore, due to the contact thermal conduction between the aluminum alloy enclosure and the COTS computing devices themselves, the impact on surface temperature and chip temperature may also be studied.

\textbf{Preliminary Results.}
In Fig. \ref{fig:ov2}, we record the surface temperature variations for 9 hours and 2 days respectively.
Fig. \ref{fig:ov2}a illustrates the variation of surface temperature over 2 days.
During 90\% of this time, the surface temperature remains at or below 4\textdegree C.
This suggests that typical COTS computing tasks exert normal influence on the surface temperature. Fig. \ref{fig:ov2}b depicts the surface temperature over a 9-hour span, a period without computing and communication tasks. In this state, the surface temperature is observed to lie within a range of 0 to 1\textdegree C. Only slight variations occur in tandem with the transition between the eclipse zone and the daylight zone.

\subsection{Chip/Surface Temperature Variations}
\label{subsec:tmp1}
\textbf{Experiment Methodology.} There are 4 principal factors influencing both the surface temperature and the chip temperature: device power consumption, computing load, task duration, and the alternating daylight and eclipse periods.
Among these, the computing load can directly affect device power consumption. To investigate these variables, we design 12 experiments. Each computing experiment is controlled to last 5 hours, with the first 3 hours (approximately equivalent to two orbital periods) devoted to computing, followed by 2 hours of halted computing, reducing the load to an idle state. For the Raspberry Pi, we measure the variations in both temperatures when utilizing 1 to 4 cores for computation (see Fig. \ref{fig:tmp1}a to Fig. \ref{fig:tmp1}d). For the Atlas, we execute image inference tasks with both single-thread and 4-thread configurations, recording the temperature variations at four power levels for each scenario (see Fig. \ref{fig:tmp1}e to Fig. \ref{fig:tmp1}i).

\textbf{Characteristic Difference.} The variation in chip temperature and surface temperature when COTS computing devices execute computing tasks exhibits significant disparities for different devices.

For the Raspberry Pi, the upper-plot on Fig. \ref{fig:tmp1} shows that they are more influenced in chip temperature by computing, with minimal effects on surface temperature. Under level 3 and level 4 workloads (see \S\ref{subsec:assumption}), the Raspberry Pi reaches a saturation temperature of approximately 82\textdegree C.
The saturation temperature refers to the point at which the COTS computing devices and satellite's external structure reach a thermal equilibrium.
This equilibrium leads to a stable thermal state. However, its surface temperature rarely exceeds 15\textdegree C, far below the operational limit, i.e. 30\textdegree C.

For the Atlas, conversely, the computing impact the surface temperature more and the chip temperature less. When Atlas operates at Full capacity with 4 threads (see Fig. \ref{fig:tmp1}i), after approximately 3 hours, the surface temperature reaches a maximum of about 24\textdegree C. This result approaches the platform's daily operational temperature design limit of 30\textdegree C. Comparative ground experiments reveal that as Atlas computational level increases from Low to Full, the power level and temperature on the satellite gradually approach the values obtained under the same computing tasks on the ground.

\textbf{Analysis.} 
The observed differences in temperature behavior between the two devices can be attributed to several factors. Firstly, the dissimilar thermal designs employ for the two devices on the \satlt\ contribute to this phenomenon (see Fig. \ref{fig:ov1}). The Atlas, with its higher power consumption, is equipped with an enhanced thermal design during the pre-launch stage. it also includes an addition of thermal conductive plates to its aluminum alloy enclosure for heat dissipation. The Raspberry Pi, in contrast, is enveloped in a simpler aluminum alloy enclosure, resulting in inferior heat dissipation.

\begin{figure*}[pt]
\centering
\captionsetup{skip=3pt, belowskip=-20pt}
\begin{minipage}{0.66\textwidth}
\includegraphics[scale=0.33, trim={0 0.2cm 0 0.1cm}, clip]{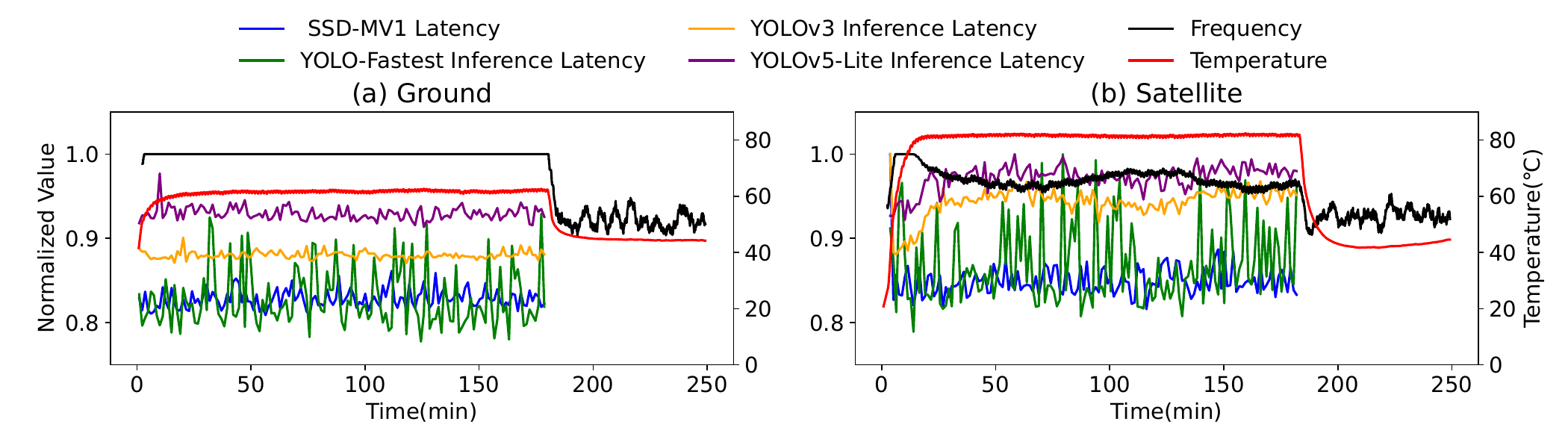}
\caption{Frequency Throttling of Pi in Space.}
\label{fig:tmp2}
\end{minipage}
\captionsetup{skip=12pt, belowskip=0pt}
\hfill
\captionsetup{skip=6pt, belowskip=-16pt}
\begin{minipage}{0.32\textwidth}
\includegraphics[scale=0.32, trim={0 0 0 0}, clip]{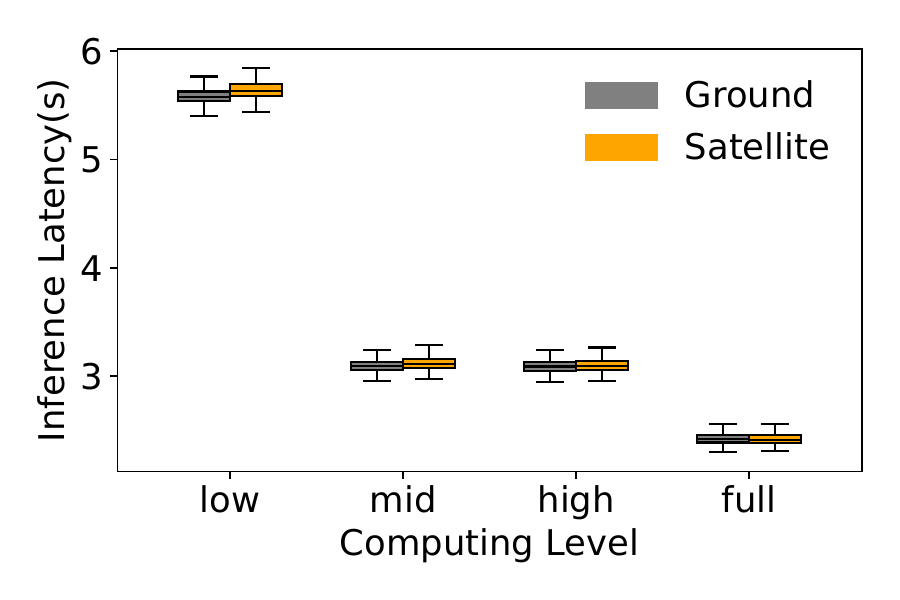}
\caption{Atlas Latency Variations.}
\label{fig:tmp3}
\end{minipage}
\captionsetup{skip=12pt, belowskip=0pt}
\Description{Void.}
\end{figure*}

\captionsetup{skip=3pt, belowskip=-16pt}
\begin{figure}[pt]
\centering
\includegraphics[scale=0.33, trim={0 0 0 0}, clip]{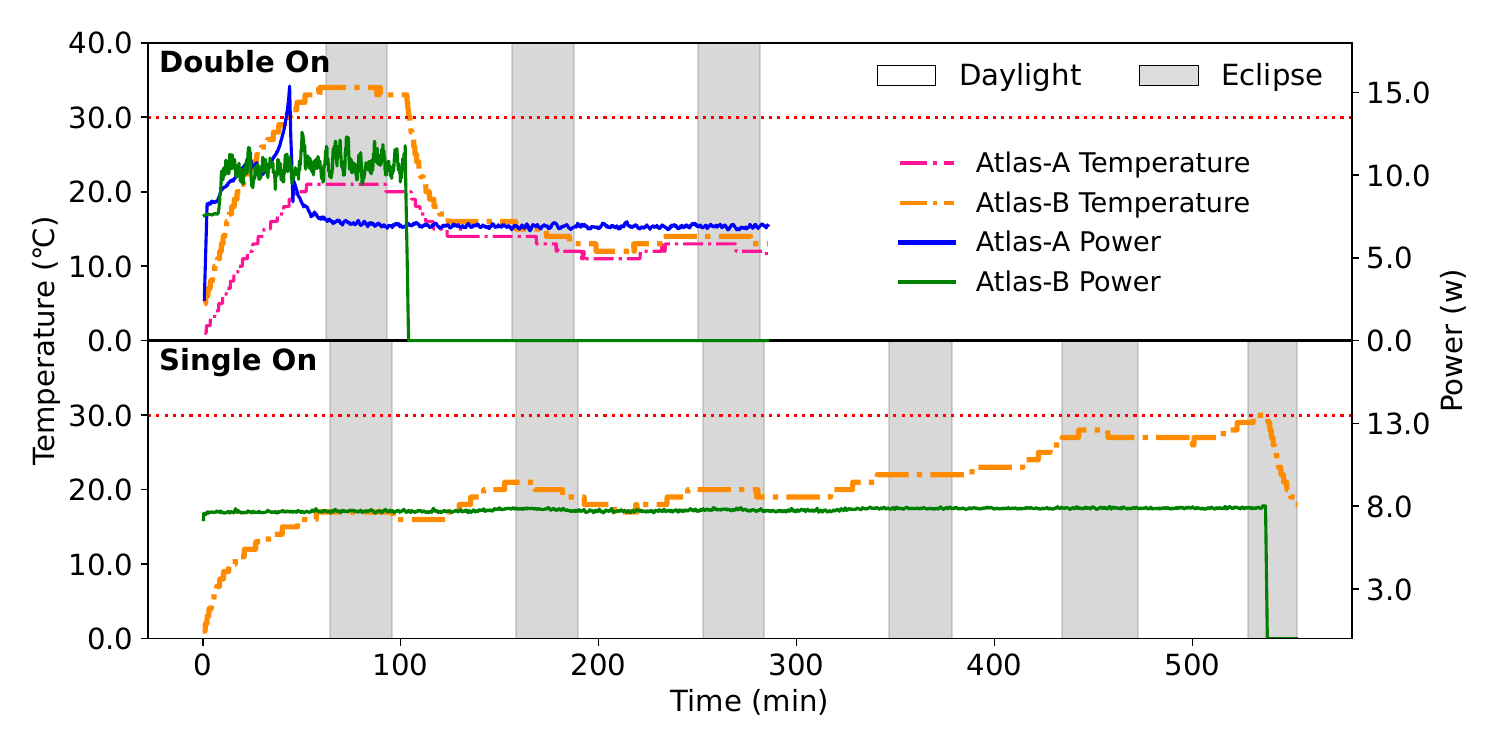}
\caption{Atlas Surface Overheating.}
\Description{Void.}
\label{fig:tmp4}
\end{figure}
\captionsetup{skip=12pt, belowskip=0pt}

Secondly, the nature of the COTS computing devices themselves plays a role. Satellite heat dissipation relies on contact cooling. The efficiency of temperature conduction depends on factors such as the thermal conductivity, thickness, and surface area of the contact material. Both the Atlas and Raspberry Pi utilize aluminum alloy and thermal pads for cooling, with similar thermal conductivity and thickness. However, they differ in chip surface area, with the Raspberry Pi's being smaller (14mm $\times$ 14mm) and the Atlas's larger (52.6mm $\times$ 38.5mm). Consequently, the reduced heat dissipation area on Raspberry Pi restricts the degree of chip cooling.
Besides these two factors, the variance in power consumption levels between the two devices may also be considered.

\textbf{Implications.} 
The experiments conducted reveal critical insights into the varying emphasis required for different COTS computing devices, particularly concerning the design of thermal structures, the scheduling of computing task duration, and the magnitude of computing loads.
(1) For low-power computing devices such as the Raspberry Pi, \textit{the utilization of materials with higher thermal conductivity} is advisable. Such an approach can slow down the time taken to reach saturation temperature and accelerate the heat dissipation process;
(2) Conversely, for high-power, high-performance devices like the Atlas, attention must be paid to \textit{controlling both the load and duration of computing tasks}. Prolonged operation must be avoided to prevent excessive surface temperature, which could potentially compromise the stability of the entire satellite platform.

\subsection{Overheating}

In the experiments in \S\ref{subsec:tmp1}, overheating phenomena occur in both the aluminum alloy enclosure and the chips: the surface temperature of Atlas approaches 30\textdegree C after 3 hours of computation; the chip temperature of Raspberry Pi escalates to above 80\textdegree C. This observation raises two pertinent questions:
\begin{list}{$\bullet$}{\leftmargin=1em \itemindent=0em \topsep=0.1em}
\item $\textbf{Q1}$: Does Atlas induce enclosure overheating with extended, high-load computing tasks?
\item $\textbf{Q2}$: What impact does elevated chip temperature have on the computing capabilities of the Raspberry Pi?
\end{list}

To investigate the \textbf{Q1}, we design two experiments. The Experiment 1 (denoted as "Single On" in Fig. \ref{fig:tmp4}) extends the task duration to 9 hours with a configuration of 4-thread and level Full on Atlas-B to monitor the surface temperature changes. The second experiment (denoted as "Double On" in  Fig. \ref{fig:tmp4}), simultaneously activates Atlas-A and Atlas-B. Atlas-B executes image inference with the same configuration in Experiment 1, and Atlas-A performs a 4-thread stress test task using only CPU. We deploy a lower load on Atlas-A, because the satellite does not support both Atlases running high-load tasks concurrently in the ground temperature testing. A target computing time of 3 hours is set for both devices.

\textbf{Enclosure Overheating.} Experiment 1 (see the lower-plot on Fig. \ref{fig:tmp4}) reveals that Atlas-B reaches approximately 17\textdegree C after 50 minutes of computation and 30\textdegree C, the operational limit, after 540 minutes. Atlas-B ceases its computing task normally at this point, and 15 minutes later, its surface temperature returns to around 17\textdegree C. Experiment 2 (see upper-plot on Fig. \ref{fig:tmp4}) indicates that Atlas-B's surface temperature reaches 30\textdegree C around 40 minutes in, persisting for about 100 minutes. Anomalously, Atlas-A is forcibly halted around 50 minutes (process killed by OS), and Atlas-B eventually gets shut down around 110 minutes. This phenomenon underscores that the operational limit of surface temperature is a key constraint on the computing duration and load of onboard COTS computing devices.

To explore the \textbf{Q2}, we deploy image inference applications using various models (see \S\ref{subsec:assumption}) on the Raspberry Pi. All the tasks employ the same task duration as in \S\ref{subsec:tmp1} to evaluate the impact of chip overheat on computing performance of the Raspberry Pi.

\textbf{Chip Overheating.} The chip overheat leads to a mild frequency throttling of approximately 5\% in the onboard Raspberry Pi. 
While an identical task running on the ground Raspberry Pi exhibits no such frequency throttling.
For Raspberry Pi, the impact of this frequency throttling varies across different models. 
In the case of YOLOv3 and YOLOv5, the average inference latency increases by roughly 10\% after the temperature reaches saturation.
Simultaneously, for inference tasks other than image segmentation, the absolute inference time rises by around 5\% to 8\%.
For Atlas, the increase in chip temperature has almost no effect on computing performance.
As the Fig. \ref{fig:tmp3} shows, the difference between satellite and ground computing performance is less than 1\%.
Furthermore, under 4 different power levels, Atlas exhibits some variation in performance. 
The Mid and High power levels improve approximately 45\% over Low.
The Full power level has around a 55\% performance boost over the Mid and High power levels.

\textbf{Analysis on Enclosure Overheating.} For the Experiment 1, \textit{the task duration} plays the key role in causing the aluminum alloy enclosure of Atlas slightly exceeds the operational limit.
In the Experiment 2, both of computing tasks on Atlas-A and Atlas-B are passively terminated due to temperature elevation.
It further illustrates that multiple devices running together intensify the rate of surface temperature increase.
This phenomenon primarily results from the thermal structure design. On \satlt, the two computing devices share an aluminum surface to conserve the internal space. Thus the computing load leads to rapid heat release when the both of the two Atlas are on.
To prevent such problem, one solution involves thermally isolating COTS computing devices. But the cost of such isolation is typically high, making it generally applicable only to critical satellite components like batteries \cite{nasa2022sotasat}.

\textbf{Analysis on Chip Overheating.} Raspberry Pi employs Dynamic Frequency Scaling (DVFS) as its CPU frequency control method. This mechanism causes Raspberry Pi to actively frequency throttling when the chip temperature exceeds 80\textdegree C. Given the cooling capabilities on \satlt, Raspberry Pi reaches an active frequency throttling zone of over 80\textdegree C within 50 minutes under load levels above 75\%. Atlas, on the other hand, does not have a frequency control mechanism similar to DVFS; instead, its power consumption level is adjusted at fixed levels across 4 power levels.

\renewcommand{\arraystretch}{0.95}
\begin{table}[t]
\centering
\caption{Temperature: Differences of Space and Ground}
\scriptsize
\begin{tabular}{|>{\centering\arraybackslash}p{2cm}|>{\centering\arraybackslash}p{1.1cm}|>{\centering\arraybackslash}p{1.1cm}|>{\centering\arraybackslash}p{1.1cm}|>{\centering\arraybackslash}p{1.1cm}|}
\hline
    & \multicolumn{2}{c|}{\textbf{Average Power (W)}} & \multicolumn{2}{c|}{\textbf{Ascending Time (min)}} \\
\cline{2-5}
    & \textbf{Space} & \textbf{Ground} & \textbf{Space} & \textbf{Ground} \\
\hline
Pi (Idle) & 2.36 & 2.80 & None& None\\
\hline
Pi (1 Core) & 3.16 & 3.63 & 161 & 67 \\
\hline
Pi (2 Core) & 3.99 & 4.47 & 155 & 41 \\
\hline
Pi (3 Core) & 4.66 & 5.26 & 125 & 38 \\
\hline
Pi (4 Core) & 4.82 & 5.92 & 86 & 39 \\
\hline
Atlas (Idle, Low) & 6.67 & 7.61 & None& None\\
\hline
Atlas (1T, Low) & 6.81 & 7.81 & 60 & 27 \\
\hline
Atlas (4T, Low) & 7.31 & 8.46 & 147 & 35 \\
\hline
Atlas (Idle, Mid) & 7.08 & 8.03 & None& None\\
\hline
Atlas (1T, Mid) & 7.35 & 8.32 & 155 & 53 \\
\hline
Atlas (4T, Mid) & 8.45 & 9.71 & 169 & 154 \\
\hline
Atlas (Idle, High) & 7.07 & 8.03 & None& None\\
\hline
Atlas (1T, High) & 7.37 & 8.42 & 148 & 40 \\
\hline
Atlas (4T, High) & 8.54 & 9.91 & 170 & 36 \\
\hline
Atlas (Idle, Full) & 7.75 & 8.63 & None& None\\
\hline
Atlas (1T, Full) & 8.17 & 9.17 & 164 & 145 \\
\hline
Atlas (4T, Full) & 10.19 & 11.48 & 147 & 88 \\
\hline
\end{tabular}
\label{tab:temp-ascending}
\end{table}
\renewcommand{\arraystretch}{1}

\textbf{Implications.}
\textit{Overheating phenomena limit the duration and load level of satellite computing and might degrade the computing performance of the SoC.}
(1) In Experiment 1, Atlas possibly reaches the 30\textdegree C operational limit in surface temperature after 9 hours of computing. This achievement occurs at an 8.5W power consumption level under the thermal conditions specific to Atlas after modification. It highlights the need for considering satellite platform temperature constraints for specific COTS computing devices under high-load computing tasks.
\textit{Excessive computing time becomes the primary limiting factor in this case.}
It necessitates the reduction of the maximum task duration for a single experiment and the setting of reasonable intervals for cooling;
(2) In Experiment 2, parallel computing with two Atlas devices leads to the structure temperature reaching the 30\textdegree C operational limit in a short time. The computing tasks on both devices are eventually forcibly terminated due to overheating. This further illustrates that onboard COTS computing devices cannot operate for extended periods under excessive load. \textit{Excessive computing load becomes the main limiting factor in this case.}
Furthermore, even without excessive load, overheating may still cause a decline in computing performance for specific devices, such as the Raspberry Pi.

The experiments in this section provide insights for \textit{modeling or conducting hardware-in-the-loop simulations} of the capabilities of onboard COTS computing devices. Computing with COTS on a satellite requires full consideration of thermal constraints. Constraints must be made on task duration, computing load, and computing performance according to the specific computing device. Additionally, it is essential to design more efficient, space-saving, and cost-effective cooling systems for the deployment of multiple COTS computing devices on the satellite.

\captionsetup{skip=3pt, belowskip=-16pt}
\begin{figure*}[pt]
\centering
\includegraphics[scale=0.37, trim={0 0 0 0}, clip]{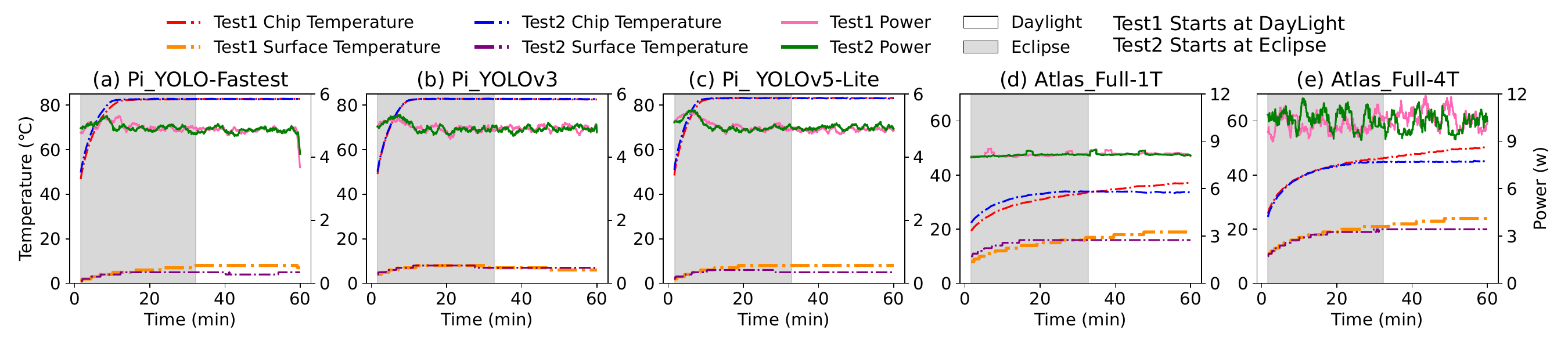}
\caption{Surface Temperature and Power Variations of COTS Computing Devices Starting at Eclipse or Daylight.}
\Description{Void.}
\label{fig:tmp6}
\end{figure*}
\captionsetup{skip=12pt, belowskip=0pt}

\begin{figure}[pt]
\centering
\captionsetup{skip=3pt, belowskip=-16pt}
\begin{minipage}{0.22\textwidth}
\includegraphics[scale=0.27, trim={0 0 0 0}, clip]{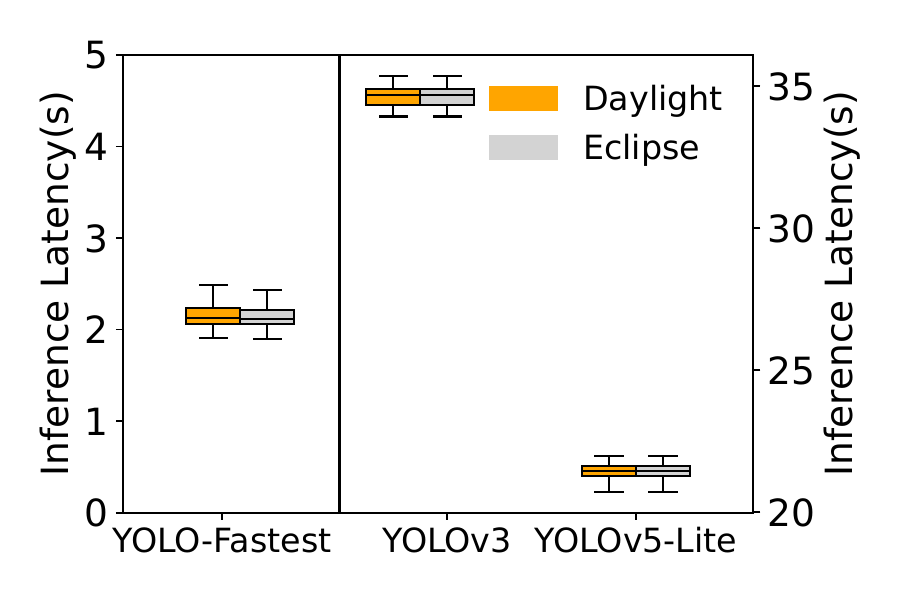}
\caption{Pi Latency.}
\label{fig:tmp5}
\end{minipage}
\captionsetup{skip=12pt, belowskip=0pt}
\hspace{0.002\textwidth}
\captionsetup{skip=3pt, belowskip=-16pt}
\begin{minipage}{0.22\textwidth}
\includegraphics[scale=0.27, trim={0.5cm 0 0.5cm 0}, clip]{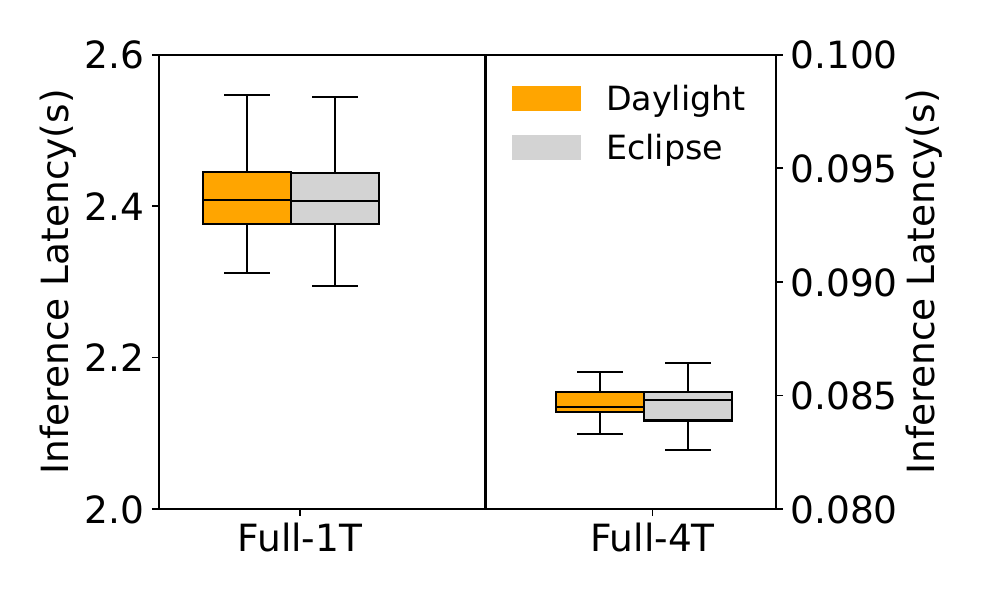}
\caption{Atlas Latency.}
\label{fig:tmp7}
\end{minipage}
\captionsetup{skip=12pt, belowskip=0pt}
\Description{Void.}
\end{figure}

\subsection{Heating Rate}
\label{subsec:tmp3}
In the experiments of \S\ref{subsec:tmp1}, differences are observed in both the power and the temperature rise rate when COTS computing devices execute identical tasks under satellite and ground conditions.
However, two problems still confront us: how to conduct a quantitative assessment of the differences in these two aspects; and how to clearly delineate the variations in the impact of space and terrestrial environments on both facets.
To clarify on these two points, we calculate the average power and times to reach maximum temperature for different computing tasks from the experiments in \S\ref{subsec:tmp1}.
Table \ref{tab:temp-ascending} illustrates the average power consumption and temperature rise time for COTS computing devices operating under different computing tasks in both satellite and ground environments. The temperature rise time quantifies the time taken by the COTS computing devices from task initiation to the attainment of maximum temperature. 

\textbf{Rate of Temperature Increase.} 
Generally, the chip temperature rises more smoothly in space than on the ground. However, for both space and ground conditions, the chip temperatures on the Raspberry Pi reach a stable state more quickly than those on the Atlas (see Fig. \ref{fig:tmp1}).
In Table \ref{tab:temp-ascending}, the time required for the satellite-based COTS computing devices to heat to saturation temperature is longer, amounting to 2-4 times the duration on the ground. Moreover, within a computing span of 3 hours, the satellite experiments in most cases reach their maximum values in the final 20\% of the time.

\textbf{Analysis.} Two factors may contribute to the disparity in heating time between satellite and ground.
First, the vast difference between the vacuum of space and ground environment matters.
Although we make sure that the both cases of space and ground utilize the same passive contact cooling mechanism (see \S\ref{subsec:3-3}).
Convection through the air allows the COTS computing devices on the ground to achieve saturation more quickly, while the vacuum in space doesn’t have the same effect \cite{nasa2022sotasat}.
Second, eclipse and daylight zones exert certain influences. for Raspberry Pi operating under level 3 and level 4 (see Fig. \ref{fig:tmp1}c and Fig. \ref{fig:tmp1}d), saturation temperature can be reached within 30 minutes. But the peak temperature is only attained after 120 minutes, largely owing to a slight temperature increase caused by passing through the daylight zone.

\textbf{Implications.} This characteristic of satellite temperature can be mitigated by intermittent computing \cite{lucia2015simpler} to reduce the average temperature over a period. This strategy leverages the slower heating rate in space, allowing for a more controlled thermal environment and potentially enhancing the overall system's reliability and performance.

\subsection{Daylight and Eclipse Zones}
From previous experiments, it is understood that the periodic variations of the daylight and eclipse zones exert a certain influence on both the surface temperature and chip temperature (see \S\ref{subsec:tmp1} and \S\ref{subsec:tmp3}).

Due to the alternating cycles of eclipse and daylight in Sun-synchronous orbits, satellite computing tasks can typically be scheduled to start either in the eclipse zone or in the daylight zone.
Thus we aim to further explore the extent to which initiating in the eclipse zone or in the daylight zone impacts satellite computing.
To elucidate this issue, we compare two scenarios for COTS computing devices: \textit{starting in the eclipse zone (denoted as Test2 in Fig. \ref{fig:tmp6}) and starting in the daylight zone (denoted as Test1 in Fig. \ref{fig:tmp6}).}
Fig. \ref{fig:tmp6} shows the two The computing tasks, lasting approximately 60 minutes, run the same computing program on both of the Raspberry Pi and Atlas.

\textbf{Impact of Eclipse and Daylight.} Figure \ref{fig:tmp6} illustrates that for the Raspberry Pi, initiating the process in either the eclipse zone or daylight zone has almost no impact on both the chip temperature and surface temperature.
However, for the Atlas, the groups starting in the daylight zone ultimately reach a temperature difference of around one hour.
In Figure \ref{fig:tmp6}, the chip temperature displays a final difference of approximately 8\textdegree C, and the surface temperature shows a difference of around 5\textdegree C.
In Fig. \ref{fig:tmp5} and Fig. \ref{fig:tmp7}, there is minimal difference in the performance of different tasks during the aforementioned process.

\textbf{Analysis.}
Solar irradiation serves as the main external heat input for the satellite.
The passive cooling structure on satellites mainly dissipates heat through its external structure \cite{tachikawa2022advanced}.
Thus the impact caused by the eclipse and daylight zones primarily results from uneven irradiation from the Sun toward the satellite external structure during its movement.
The difference in thermal conductivity remains the main cause of the distinct temperature performances between these two devices.

\textbf{Implications.}
(1) The differences in starting within the daylight or eclipse zone have minimal impact on computing performance.
This finding implies that for hardware-in-the-loop simulations targeting satellite computing, this effect can be simplified;
(2) On the other hand, in the scheduling of satellite computing tasks, the effect of computing in the eclipse zone or first in the daylight zone on temperature tends to become consistent as the total time increases.
\section{Energy Results}

\captionsetup{skip=3pt, belowskip=-16pt}
\begin{figure}[t]
\centering
\includegraphics[scale=0.255, trim={0 0.8cm 0 0}, clip]{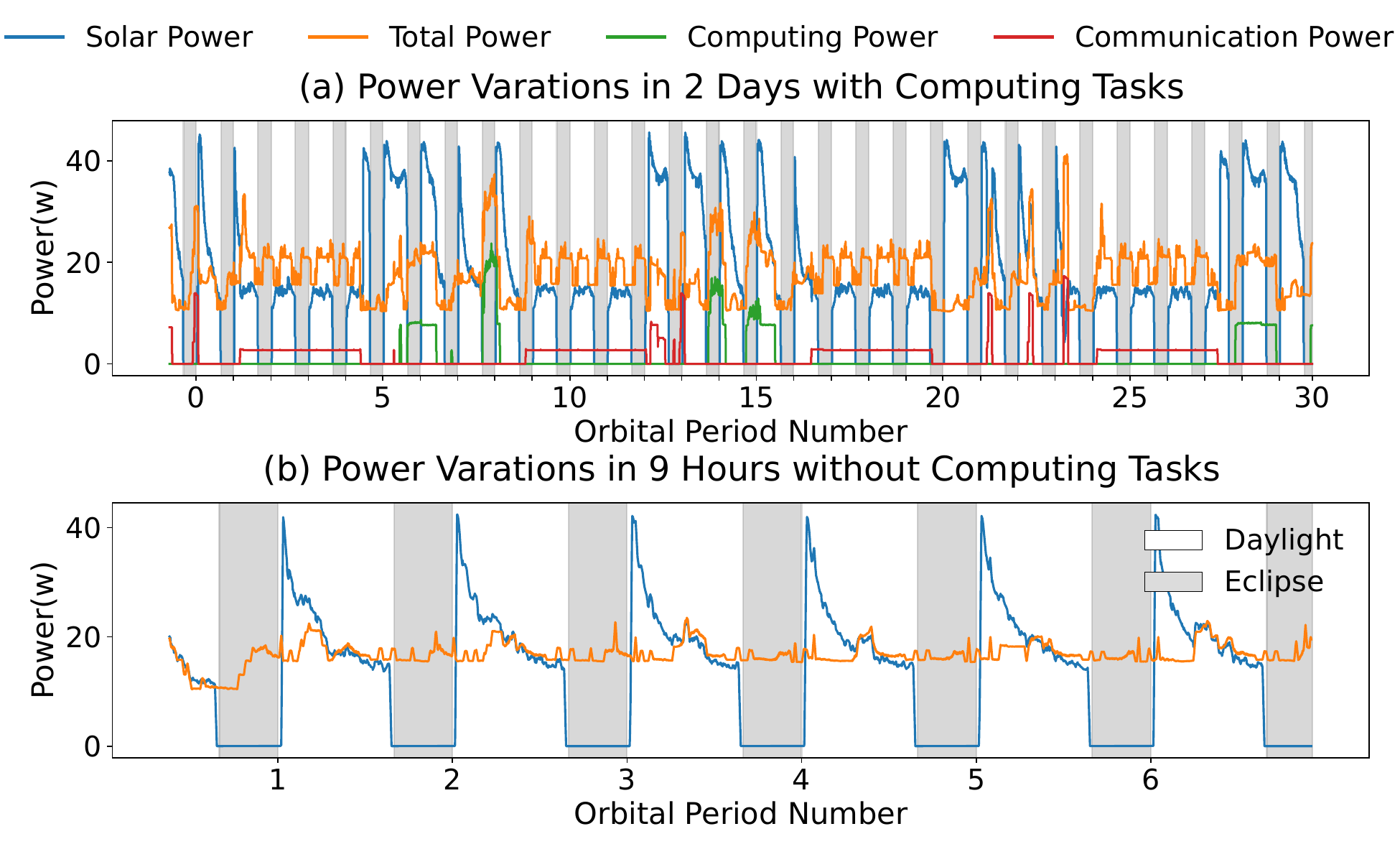}
\caption{Power Overview.}
\Description{Void.}
\label{fig:ov3}
\end{figure}
\captionsetup{skip=12pt, belowskip=0pt}

\begin{figure*}[pt]
\centering
\captionsetup{skip=3pt, belowskip=-16pt}
\begin{minipage}{0.3\textwidth}
\includegraphics[scale=0.22, trim={0 0 0 0}, clip]{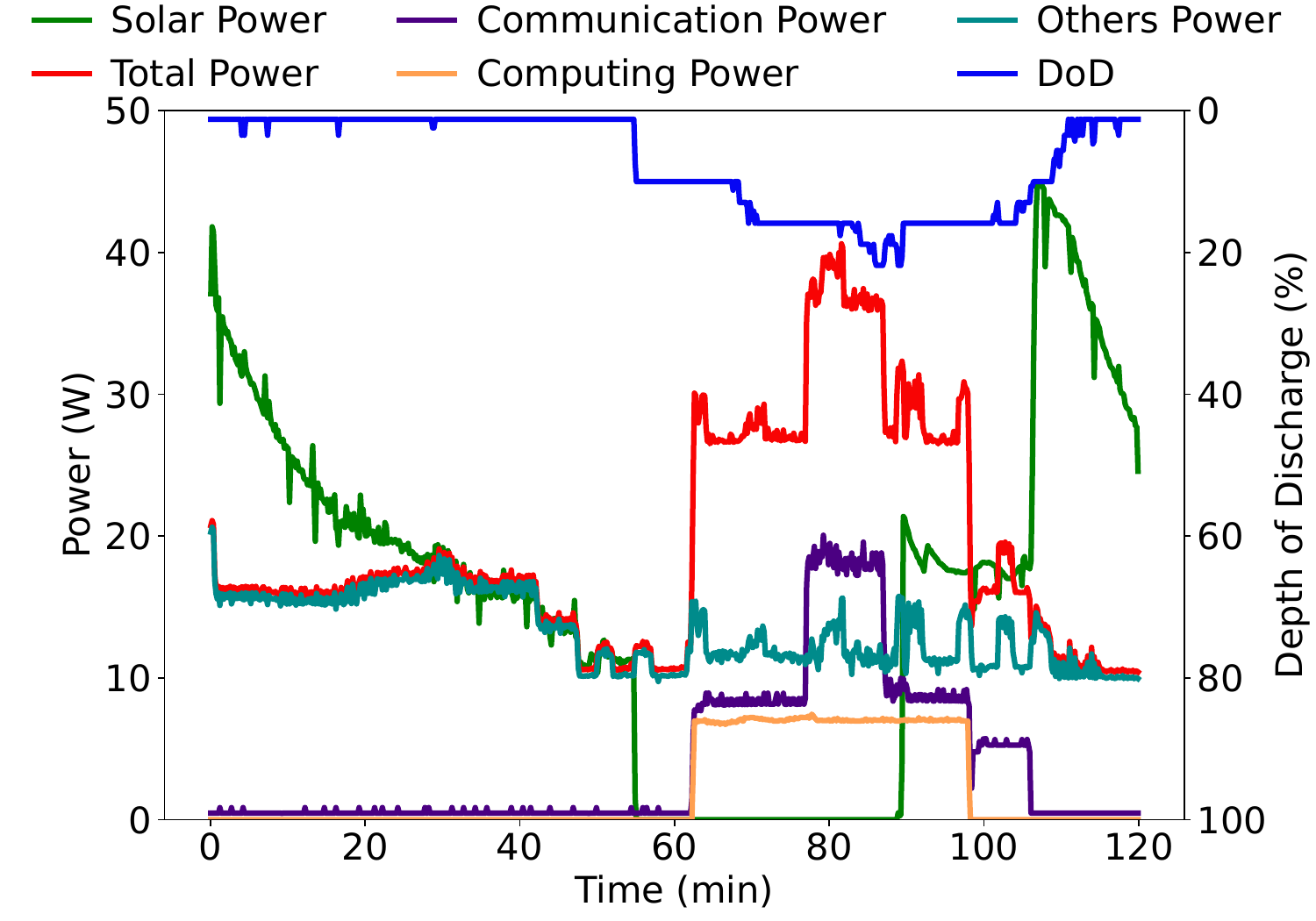}
\caption{Power Variation.}
\label{fig:pwr1}
\end{minipage}
\captionsetup{skip=12pt, belowskip=0pt}
\hspace{0.002\textwidth}
\captionsetup{skip=3pt, belowskip=-16pt}
\begin{minipage}{0.33\textwidth}
\includegraphics[scale=0.22, trim={1cm 0 1.5cm 1cm}, clip]{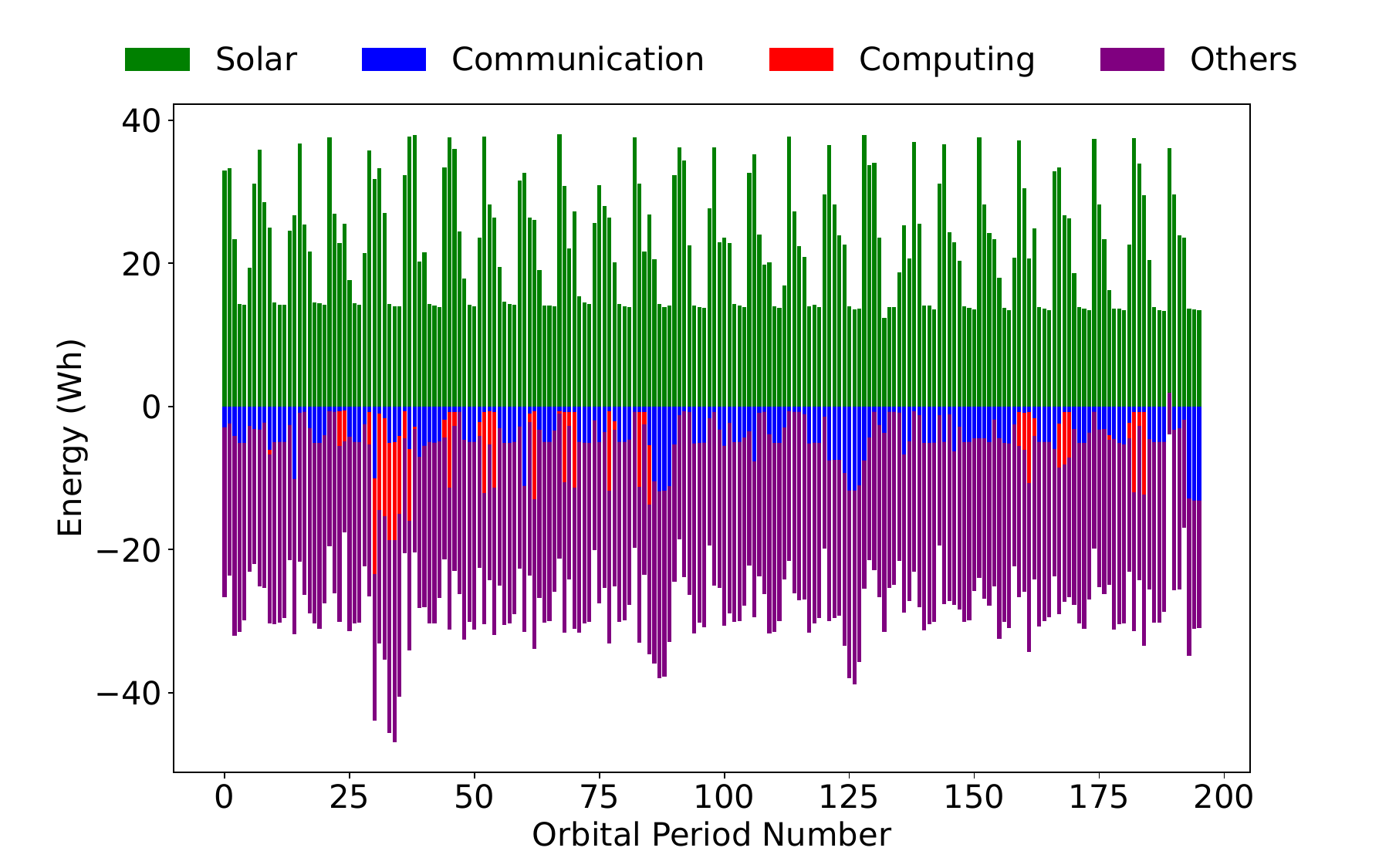}
\caption{Energy Variation per Orbit.}
\label{fig:pwr2}
\end{minipage}
\captionsetup{skip=12pt, belowskip=0pt}
\hspace{0.000\textwidth}
\captionsetup{skip=3pt, belowskip=-16pt}
\begin{minipage}{0.33\textwidth}
\includegraphics[scale=0.22, trim={1cm 0 2cm 1cm}, clip]{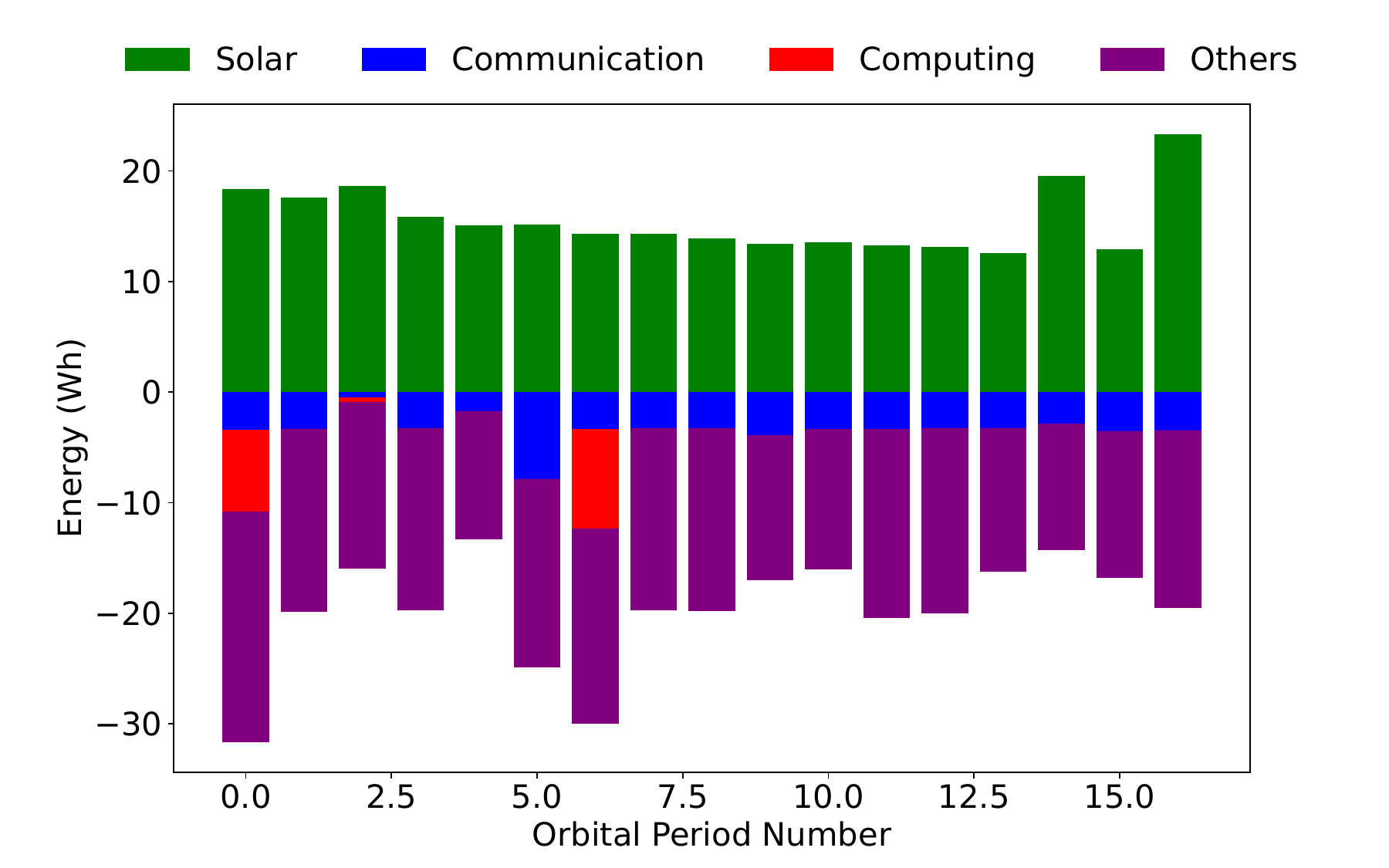}
\caption{Energy Variation per Day.}
\label{fig:pwr3}
\end{minipage}
\captionsetup{skip=12pt, belowskip=0pt}
\Description{Void.}
\end{figure*}

The Electrical Power Subsystem (EPS) is responsible for handling power generation, storage, and distribution within LEO satellites. The satellite energy recycling relies primarily on solar panels \cite{socolovsky2017development}, rechargeable lithium batteries \cite{fellner2003lithium}, and power distribution/control board. As of 2021, approximately 85\% of nanosatellite power systems apply these setups \cite{UCS}. The EPS efficiently oversees energy conversion, distribution, and regulation, while preserving battery life and catering to the rigors of space and the fluctuating power needs of various mission phases \cite{JAXA-JERG-2-214, nasa2022sotasat}.

In terms of energy collection, the deciding factor is the periodic change in sunlight conditions caused by satellite motion. Regarding energy usage, we concentrate on the power consumption of COTS computing devices.
As seen in the preliminary measurement results in Fig. \ref{fig:ov3}, computing tasks occupy a significant portion of the non-routine energy consumption. For energy storage, the DoD is a key index determining satellite life. We aim to control the \satlt's DoD level below 30\%. The general design target is approximately 25\% DoD\cite{wertz1999space}, so 30\% is not considered excessive.

\label{subsec:power}
\textbf{Preliminary Results.}
The variation of harvested solar power and individual power changes in computing and communication devices over two days are shown in Fig. \ref{fig:ov3}a. Power consumption pattern reveals four cycles in a day where the average power consumption reaches 40W, with other times approximately 20W. We make further study about this in \S\ref{subsec:energy-harv}.
Fig. \ref{fig:ov3}b presents power changes over 10 hours, a period devoid of computing and communication tasks, with the corresponding power consumption being zero. The harvested solar power and the total power consumption of the satellite platform maintain dynamic balance.

\textbf{Metrics.}
In Fig. \ref{fig:pwr1}, the \textit{Solar Power} refers to the power generated by MPPT (see \S\ref{subsec:3-3}) to be used to charge the battery and supply COTS computing devices or other components.
The \textit{Total Power} refers to the power consumed by the satellite, including all kinds of consumed power.
The \textit{Communication Power}, \textit{Computing Power}, and \textit{Others Power} refers to the consumed  power by the communication components, COTS computing devices, and other components on the satellite platform.
In Fig. \ref{fig:pwr2} and Fig. \ref{fig:pwr3}, we use the corresponding terms, i.e. \textit{Solar}, \textit{Communication}, \textit{Computing}, and \textit{Others}, to denote the energy generated or consumed by the corresponding components.

\subsection{Energy Harvesting Patterns}
\label{subsec:energy-harv}

\textbf{Short-term Power Variation.} Figure \ref{fig:pwr1} analyzes a typical cycle featuring a data transmission task.
At around 60 minutes, communication power rises to about 18W due to the activation of communication components. At the same time, computing power rises for the activation of Atlas-B. At around 78 minutes, the peak is caused by the beginning of the data transmission.
The solar power reaches a peak of approximately 40W within about 20 minutes. It subsequently decline to around 10W, before finally entering the shadowed region and reducing to zero. During the initial 30 minutes depicted in Figure \ref{fig:pwr1}, the batteries are nearly fully charged. The solar power exceeds the total power of the satellite platform, resulting in about 3Wh of the solar energy not being utilized by the satellite.

\textbf{Periodic Energy Variation.} 
In Fig. \ref{fig:pwr2}, we adjusted the granularity of time, observing energy variation on the orbital periods. 
It can be observed that \satlt\ exhibits periodicity in energy collection at the granularity of an orbital period. Approximately every 15 orbital periods, corresponding to roughly 24 hours, there are two peaks and valleys depicted in Fig. \ref{fig:pwr2}. This pattern reveals that the power generation capacity of \satlt's solar panels initiates a new cycle approximately every 24 hours.

In Fig. \ref{fig:pwr3}, we observe over a specific 60-min time interval in daylight per day.
Despite the cyclical nature of energy harvesting, the energy yield remains relatively stable. Within each round, the energy collected by the solar panels predominantly clusters around the average value of 16Wh, without manifesting pronounced peaks and valleys.

\textbf{Analysis.} The satellite's energy collection shows regular patterns over two different time scales: the orbital period and 24 hours. When the cycle is defined by the orbital period, the periodic variation in energy collection primarily stems from the relative motion between the satellite and the Sun, leading to periodic changes in sunlight exposure. In the eclipse zone, the solar irradiance is zero, while in the daylight zone, the satellite's power undergoes an initial increase followed by a decrease. This pattern results mainly from the alteration in the angle of the satellite's solar panels during motion, thereby inducing corresponding fluctuations in the energy received by the solar panels.

When the cycle spans 24 hours, the formation of daily energy collection cycles is attributable to the working mode of the EPS. The EPS can be configured with a power profile to augment the energy reception of the solar panels at specific intervals\cite{wertz1999space}. However, this enhancement is not without constraints. An increase in the power of the solar panels would elevate their temperature, which must not surpass the tolerance threshold.
The continuous cycles of collected energy depicted in Fig. \ref{fig:pwr2} are illustrative of \satlt's EPS working mode. Out of approximately 15 cycles a day, 10 exceed 14Wh, while the remaining five cycles hover around 14Wh.

\textbf{Implications.} (1) \textit{Stability and Predictability of Energy Collection:} The periodic nature of the satellite's energy collection ensures that the acquired energy is both stable and predictable. From the perspective of factors determining satellite energy collection across two time dimensions, they demonstrate relative stability over longer time cycles. On one hand, the satellite's orbital changes and its ability to harness solar energy can be predicted within a small margin of error; on the other hand, the working mode of the EPS is fundamentally determined at the time of the satellite's fabrication, so its impact on energy harvesting is also well predictable; (2) \textit{Maximizing Utilization of Collected Energy with Satellite Computing:} As depicted in Fig. \ref{fig:pwr1}, the energy collected by the satellite is not fully utilized. Given the characteristics of satellite computing (see \S\ref{subsec:assumption}), employing the unutilized energy to perform computing tasks helps processing the excessive data. The predictability of solar power can also help the scheduling of computing tasks to avoid energy under-utilization.

\subsection{Energy Expenditure Breakdown}

\textbf{Short-term Power Variation.} In Fig. \ref{fig:pwr1}, at around the 62nd minute, the satellite activates the computing and communication devices, preparing for data transmission. At around the 78th minute, the satellite commences the data transmission, lasting for around 10 minutes, with the communication power staying 38W. Meanwhile, the power consumption of the COTS computing device remains around 14W throughout. Overall, short-term communication and computation power consumption have a significant impact on the battery's DoD, consuming 20\% within 30 minutes.

\textbf{Periodic Energy Variation.} In Fig. \ref{fig:pwr2}, the energy consumption of communication remains relatively steady, basically maintaining around 5Wh per round most of the time. But the short-term share of computing energy can even reach 30\% to 40\%. From Fig. \ref{fig:pwr3}, energy collection and consumption maintain balance in 24-hour intervals, with absolute differences generally not exceeding 20\%.

\begin{figure}[pt]
\centering
\includegraphics[scale=0.33, trim={0 0.5cm 0 0}, clip]{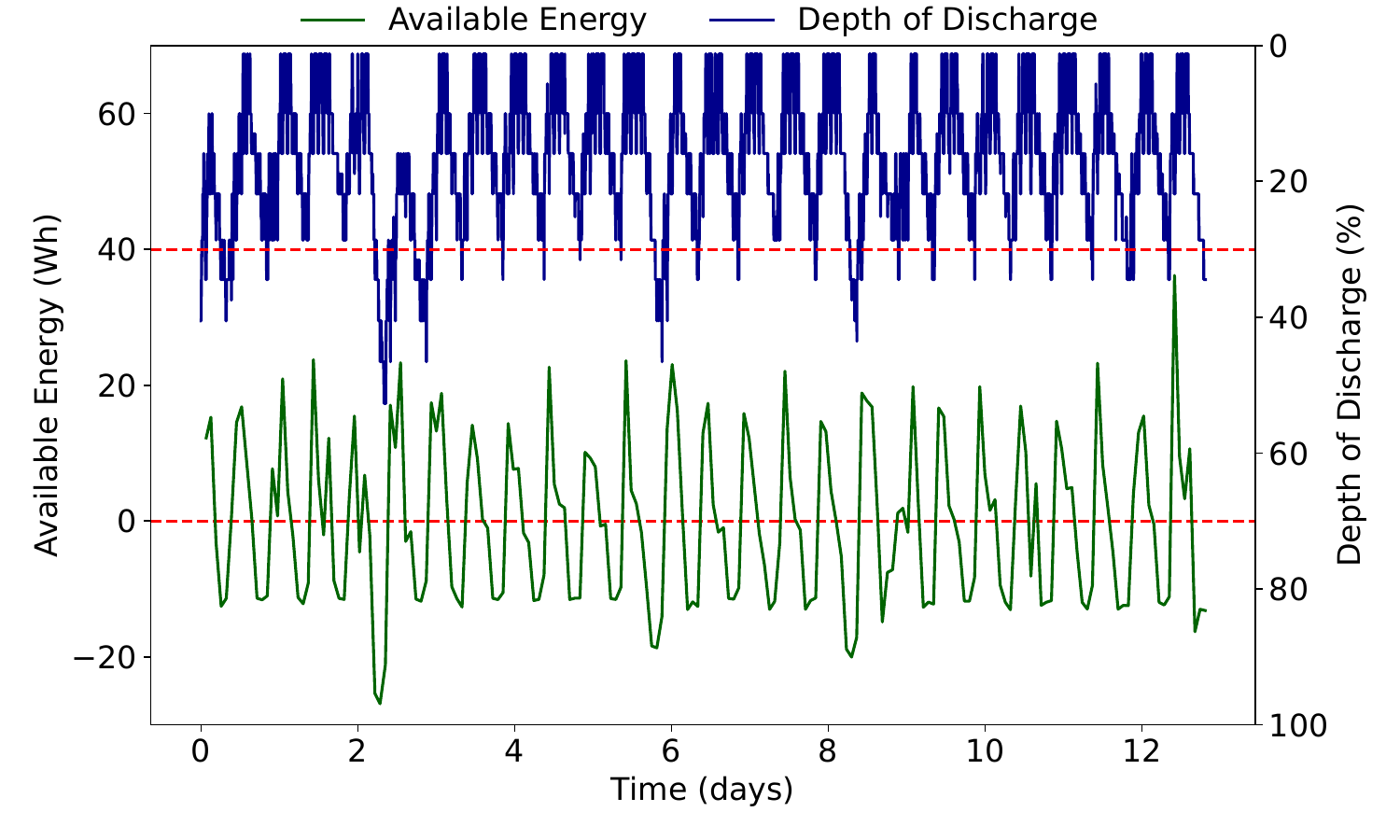}
\caption{DoD and Available Energy Variation.}
\Description{Void.}
\label{fig:pwr4}
\end{figure}

\textbf{Depth of Discharge and Available Energy.} 
Fig. \ref{fig:pwr4} illustrates the variations in DoD and the available energy for each cycle, corresponding to the same time frame shown in Fig. \ref{fig:pwr2}.
The available energy is calculated by subtracting the energy consumed from the energy collected.
The figure also marks the 30\% DoD limit and the equilibrium situation where available energy is 0. During the two-week testing period, the satellite maintains an average 16\% DoD level. However, during the time period from the second to the third day, the average DoD rises to around 35\%, which can severely affect battery degradation.
Within a short timeframe, the DoD can even exceeded 50\%.
From the perspective of available energy, the expected value is 0, representing energy balance within a cycle. But on one hand, there is still 6\% excess energy converted from solar energy that is not utilized when the battery is full. On the other hand, there is insufficient energy during compute-intensive tasks.

\textbf{Analysis.}
Communication power can rise rapidly in a short time.
On a broader time scale, the communication power is more gradual. This is mainly because communication time is relatively limited.
Meanwhile, computing energy consumption can greatly exceed the energy collected by the solar panels as shown in Fig. \ref{fig:pwr2}. Fig. \ref{fig:pwr3} shows the share of computing task energy consumption remains high. 
According to the information in Fig. \ref{fig:pwr2} for the corresponding time frame, the COTS device is performing continuous computing tasks from the second to the third day. It results in the DoD level exceeding the 30\% limit during that time frame as shown in Fig. \ref{fig:pwr4}.
Additionally, the consumed energy other than communication and computing is relatively stable.

\textbf{Implications.} (1) \textit{Dominance of Computing Energy Consumption:} The computing tasks are the primary consumers of the variable energy, considering both of power magnitude and duration. Within an orbital period, communication tasks dominate most of the power consumption if present. But over extended time frames exceeding a day, the proportion of energy consumed by communication is not substantial;
(2) \textit{Necessity of Reasonable Scheduling for Satellite Computing:}
Onboard COTS computing devices must carefully schedule computing tasks to fully utilize energy and to avoid exceeding the DoD limit.

\captionsetup{skip=13pt, belowskip=-16pt}
\captionsetup{skip=12pt, belowskip=0pt}

\subsection{Computational Energy Efficiency}
The power supply for onboard COTS computing devices is exclusively provided by the satellite platform. To investigate whether computing efficiency in space differs from that on Earth, we have deployed various types of computing tasks (see \S\ref{subsec:assumption}) both on the satellite and on the ground. The satellite is powered by solar panels and batteries, while the ground uses a programmable power monitor. This comparative study aims to uncover potential disparities and understand the underlying factors that might contribute to any observed differences in computing performance.

\textbf{Comparisons.} Tables \ref{tab:pwr-consump-atlas} and \ref{tab:pwr-consump-pi} present the energy consumption for processing 100 images on Atlas and Raspberry Pi, respectively, as well as the number of images processed per 60-minute.
From the tables, we can conclude that: 
(1) Computing tasks on the satellite are more energy-efficient than those on the ground. For the same task on Raspberry Pi, the difference in energy consumption between the ground and space is trivial (around 2\%). For the same task on Atlas, the difference is about 10\%;
(2) Computing tasks on the ground are more performant than those on satellites. When comparing the number of inference tasks, the ground is greater than or equal to the space. But this is at the cost of more power consumed (see Table \ref{tab:temp-ascending}) in the same time frame;
(3) As the computing load increases, the efficiency of energy utilization becomes higher.
Tables \ref{tab:pwr-consump-atlas} shows that when Atlas is at the Full level, the energy consumption for processing 100 images is the least.

\newcommand{\splitcell}[2][c]{
\begin{tabular}[#1]{@{}c@{}}#2\end{tabular}}
\begin{table}[ht]
\renewcommand{\arraystretch}{0.95}
\centering
\footnotesize
\caption{Energy Consumption on Atlas.}
\begin{tabular}{|>{\centering\arraybackslash}p{0.65cm}|>{\centering\arraybackslash}p{1cm}|>{\centering\arraybackslash}p{1.1cm}|>{\centering\arraybackslash}p{1.1cm}|>{\centering\arraybackslash}p{1.1cm}|>{\centering\arraybackslash}p{1.1cm}|}
\hline
\textbf{Task} & \textbf{Level} & \textbf{Sat Energy (J)} &\textbf{Gnd Energy (J)} & \textbf{Sat Quantity} & \textbf{Gnd Quantity} \\
\hline
\multirow{8}{*}{od} & 1T, Low & 3983.7 & 4407.3 & 1846 &1912\\
\cline{2-6}
& 1T, Mid & 2448.2 & 2613.8 & 3242 & 3439\\
\cline{2-6}
& 1T, High & 2428.2 & 2644.2 & 3279 & 3440\\
\cline{2-6}
& 1T, Full & 2151.7 & 2281.1 & 4098 & 4342\\
\cline{2-6}
& 4T, Low & 180.4 & 203.6 & 33 & 34\\
\cline{2-6}
& 4T, Mid & 103.3 & 117.4 & 67 & 68\\
\cline{2-6}
& 4T, High & 104.2 & 120 & 68 & 68\\
\cline{2-6}
& 4T, Full & 94.9 & 106.2 & 89 & 89\\
\hline
\multirow{2}{*}{imgcl} & Mid & 1138 & 1173 & 2310 & 2641\\
\cline{2-6}
    & Full & 1024 & 1023& 2832 & 3319\\
\hline
\multirow{2}{*}{jpege} & Mid& 645 & 665& 4030 & 4601\\
\cline{2-6}
    & Full & 550 & 566& 5207 & 5896\\
\hline
\end{tabular}
\label{tab:pwr-consump-atlas}
\end{table}

\begin{table}[ht]
\centering
\footnotesize
\renewcommand{\arraystretch}{0.95}
\caption{Energy Consumption on Pi.}
\begin{tabular}{|>{\centering\arraybackslash}p{1.4cm}|>{\centering\arraybackslash}p{1.3cm}|>{\centering\arraybackslash}p{1.3cm}|>{\centering\arraybackslash}p{1.3cm}|>{\centering\arraybackslash}p{1.3cm}|}
\hline
\textbf{Models} & \textbf{Sat Energy (J)} &\textbf{Gnd Energy (J)} & \textbf{Sat Quantity} & \textbf{Gnd Quantity} \\
\hline
yolo-fastest & 21 & 21.38 & 1640 & 1742\\
\cline{1-2} \cline{2-5}
yolo-v5lite & 207.6 & 212.9 & 168 & 176\\
\cline{1-2} \cline{2-5}
yolov3 & 335 & 345.2 &105 & 113\\
\hline
\end{tabular}
\label{tab:pwr-consump-pi}
\end{table}

\textbf{Analysis.} Conclusions (1) and (2) share a common root cause. This root cause pertains to the instability in the power supply of COTS computing devices onboard satellites, leading to a slight decline and consequently a lower overall power profile.
However, under these reduced power levels, the performance degradation of COTS computing devices remains minimal.
Two factors contribute to this outcome.
First, in the daylight zone, the satellite's attitude affects the stability of the solar power.
Second, in the eclipse zone, the voltage of batteries may decline due to either an increase in the DoD or battery aging.
Conclusion (3) aligns with intuitive reasoning.
As the computational load escalates, the additional power is exclusively channeled towards computing, hence, under higher computational load, computing efficiency improves.

\textbf{Implications.} (1) \textit{On hardware-in-the-loop simulations:} Overall, the computing efficiency under satellite power conditions proves superior. This insight implies that terrestrial hardware-in-the-loop simulations may overestimate the energy measurements, while the actual satellite computing efficiency likely experiences a moderate increase; (2) \textit{On satellite computing scheduling:} Given that higher loads yield superior efficiency ratios, COTS computing devices should ideally operate under elevated loads. However, this high-load scenario elevates the risk of thermal constraints, necessitating a comprehensive consideration of computing limitations during task scheduling.

\section{Related Work}

\textbf{Satellite System and Measurement.} Owing to advancements in rocket technology and the deployment of COTS computing devices \cite{nepp2021processor,nepp2021esa}, the costs of LEO satellite systems have significantly reduced \cite{jones2018recent}. In recent years, an emerging trend centers around system and measurement works grounded in real satellites and authentic data. These efforts can be broadly classified into the following categories:
(1) \textit{Real Satellite Systems} \cite{narayana2020hummingbird, DBLP:conf/mobicom/SinghPZYK21, singla2021satnetlab, wang2021tiansuan};
(2) \textit{Satellite System Measurements} \cite{perdices2022satellite, michel2022first, kassem2022browser};
(3) \textit{Satellite Systems Based on Authentic Data} \cite{vasisht2020distributed, rodrigues2019extracting, DBLP:conf/sigcomm/VasishtSC21}. 
The majority of existing work focuses on integrated terrestrial-satellite Internet systems \cite{DBLP:conf/mobicom/SinghPZYK21, singla2021satnetlab, perdices2022satellite, michel2022first, kassem2022browser, vasisht2020distributed, DBLP:conf/sigcomm/VasishtSC21}.
While a minor fraction is devoted to specific computing tasks \cite{wang2021tiansuan} or applications \cite{rodrigues2019extracting, narayana2020hummingbird}.
The trend toward the convergence of satellite system networking and computing is becoming increasingly pronounced, facilitating the emergence of new practical application scenarios \cite{DBLP:conf/hotnets/BhattacherjeeKL20, DBLP:conf/infocom/LaiLWLXW22}.

\textbf{Orbital Edge Computing.}
Due to the rapid data generation on satellites, which significantly outpaces the satellites' data transmission capabilities \cite{nasa2015esds, doug2020teraspace}, Onboard Edge Computing (OEC) \cite{DBLP:conf/hotnets/BhattacherjeeKL20, denby2019orbital, DBLP:conf/asplos/DenbyL20, lucia2021computational, DBLP:conf/asplos/DenbyCCLN23} uses an architecture for executing in-orbit computing tasks using COTS devices. The prevailing demand for in-orbit processing emanates primarily from the remote sensing domain \cite{chi2016big}.
Through in-orbit processing, valuable insights are swiftly and real-time extracted from voluminous remote sensing imagery, or conversely, irrelevant data are discarded. In this context, a multitude of novel algorithms and models have emerged \cite{reichstein2019deep, kucik2021investigating, SmartSat2021, uzkent2020efficient, wang2023first}, underpinned by COTS computing platforms and massive image datasets.
Concurrently, limitations such as energy consumption of COTS computing platforms receive increasing attention within the OEC architecture \cite{DBLP:conf/asplos/DenbyL20, DBLP:conf/asplos/DenbyCCLN23, li2021towards, wang2023first}. Additionally, as the scenarios and requirements for in-orbit processing proliferate  \cite{DBLP:conf/hotnets/BhattacherjeeKL20, DBLP:conf/infocom/LaiLWLXW22}, task scheduling in in-orbit computing gains in significance \cite{li2021towards, li2021service, wang2022enhancing, wang2018multi}.

\textbf{Testbed for Satellite Research.}
Historically constrained by the high costs and complexity of real systems, research in the domain of satellite networking and computing primarily relies on theoretical modeling and simulation platforms \cite{wang2022enhancing, DBLP:conf/hotnets/HauriBGS20, DBLP:conf/infocom/LaiLWLXW22, DBLP:conf/middleware/PfandzelterB22, DBLP:conf/hotedge/BhosaleBG20, DBLP:conf/hotnets/BhattacherjeeKL20, DBLP:conf/asplos/DenbyL20, kassing2020exploring, lai2020starperf}.
Significant advancements have occurred in theoretical analysis \cite{li2021service, chen2021mobility}. But a lack of in-depth system work and empirical validation in real-world environments often results in conclusions that diverge considerably from actual scenarios.
This growing divergence underscores the increasing emphasis on the development of hardware-in-the-loop simulation platforms \cite{lai2023starrynet} and even fully physical experimental setups \cite{singla2021satnetlab, wang2021tiansuan, o2023extension}.
Platforms like StarryNet \cite{lai2023starrynet} utilize a hybrid simulation approach to offer the advantage of simulation flexibility while partially maintaining system realism and credibility.
Fully physical experimental platforms require substantial initial investment but excel in reproducing the challenges and issues inherent to real-world scenarios.
As the costs of satellite manufacturing and launching continue to decline, we envision an increasing proportion of physical components in more experimental platforms \cite{lai2023starrynet, singla2021satnetlab, wang2021tiansuan, o2023extension}.

\section{Discussion on Radiation Effects}
\label{subsec:discus}

In the context of space environmental effects on spacecraft, radiation effects refer to the energy transfer by charged particles \cite{wertz1999space}. The conduction or ionization caused by a charged particle can produce various effects on semiconductor structures, with Single Event Effects (SEE) being a common category \cite{mazur2011timescale}, including Single Event Upsets (SEU) and Single Event Latchups (SEL) \cite{siegle2015mitigation}. SEUs are errors caused by bit flips in memory or registers during computation. Specifically, the affected area may contain data or instructions, and such errors can be detected and corrected at the software level. SEL represents a more severe error, such as a sudden short circuit in a charged transistor, where prolonged high current can lead to device burnout, causing irreversible damage to the spacecraft.

Protection against SEEs involves multiple dimensions, including reliability designs such as Triple Modular Redundancy (TMR) \cite{glein2015reliability} and Byzantine algorithms \cite{veronese2011efficient} at both software and hardware levels. Through integrated software and hardware system design \cite{yuan2022towards}, terrestrial high-performance COTS computing devices are also expected to meet aerospace-grade reliability standards.

Due to the protective effect of the Earth's magnetic field, SEEs are less pronounced in low Earth orbits. In our 6-month computational experiment, no computational errors caused by SEEs were observed, despite our computing devices not undergoing special radiation hardening. This demonstrates that COTS computing devices selected with simple electromechanical and thermal modifications are broadly reliable in addressing radiation effects at LEO around 500KM.
\section{Conclusion}

To conclude, in this paper, we build a real satellite system to study the impact of the thermal control and power management on the COTS computing devices in terms of computing constrains and scheduling for computing tasks. We will release the comprehensive datasets and hope to help the community make further progress on the research of the satellite computing system.

\begin{acks}
We sincerely thank our shepherd and all anonymous reviewers for their valuable feedback.
This work was supported by National Natural Science Foundation of China under Grant (NSFC No. 62032003, U21B2016, 62302055, and 62302015) and Postdoctoral Innovation Talents Support Program (No.BX20230011).
Ao Zhou is the corresponding author of this work.
\end{acks}

\bibliographystyle{ACM-Reference-Format}
\bibliography{ref}

% \appendix

\end{document}